\documentclass[10pt]{article}

\usepackage{amsmath}
\usepackage{amssymb}

\usepackage{graphicx}

\usepackage{cite}

\usepackage{color}

\usepackage[labelfont=bf,labelsep=period,justification=raggedright]{caption}

\makeatletter
\renewcommand{\@biblabel}[1]{\quad#1.}
\makeatother

\date{}

\begin{document}

\begin{flushleft}
{\Large
\textbf{A computational study of the effects of remodelled electrophysiology and mechanics on initiation of ventricular fibrillation in human heart failure} 
}
\\
\vspace{\baselineskip}
Nathan R Kirk$^{1}$, Alan P Benson$^{2}$, Christopher E Goodyer$^{1,3}$, Matthew E Hubbard$^{1\ast}$ \\
\bf{1} School of Computing, University of Leeds, Leeds, West Yorkshire, LS2 9JT, United Kingdom \\
\bf{2} School of Biomedical Sciences, University of Leeds, Leeds, West Yorkshire, LS2 9JT, United Kingdom \\
\bf{3} The Numerical Algorithms Group, Wilkinson House, Jordan Hill Road, Oxford, OX2 8DR, United Kingdom \\
$\ast$ E-mail: m.e.hubbard@leeds.ac.uk
\end{flushleft}

\section*{Abstract}

 The study of pathological cardiac conditions such as arrhythmias, a major cause
of mortality in heart failure, is becoming increasingly informed by computational
simulation, numerically modelling the governing equations. This can provide insight
where experimental work is constrained by technical limitations and/or ethical
issues. It is now commonly used to provide comprehensive studies of the mechanisms
underlying the onset of arrhythmias at cell, tissue and organ levels, based on
increasingly detailed descriptions of the electrophysiology, derived from experiment.

 As the models become more realistic, the construction of efficient and accurate
computational models becomes increasingly challenging. In particular, recent
developments have started to couple the electrophysiology models with mechanical
models in order to investigate the effect of tissue deformation on arrhythmogenesis,
thus introducing an element of nonlinearity into the mathematical representation.
This paper outlines a biophysically-detailed computational model of coupled
electromechanical cardiac activity which uses the finite element method to
approximate both electrical and mechanical systems on unstructured, deforming,
meshes. An ILU preconditioner is applied to improve performance of the solver.

 This software is used to examine the role of electrophysiology, fibrosis and
mechanical deformation on the stability of spiral wave dynamics in human ventricular
tissue by applying it to models of both healthy and failing tissue. The latter was
simulated by modifying (i) cellular electrophysiological properties, to generate an
increased action potential duration and altered intracellular calcium handling, and
(ii) tissue-level properties, to simulate the gap junction remodelling, fibrosis
and increased tissue stiffness seen in heart failure. The resulting numerical
experiments suggest that, for the chosen mathematical models of electrophysiology
and mechanical response, introducing tissue level fibrosis can have a destabilising
effect on the dynamics, while the net effect of the electrophysiological remodelling
stabilises the system.


\section*{Introduction}

 Heart failure accounts for over 280000 deaths each year in the United States
alone \cite{AHA2010}, with up to 50\% of these deaths being sudden and presumably
due to ventricular arrhythmia \cite{Tomaselli2004}. Ventricular fibrillation is
an often fatal arrhythmia, in which the normal sinus rhythm is disturbed when
the same wavefronts continually re-excite the same tissue (re-entry); synchronous
contraction of the ventricles is lost, circulation of the blood ceases and death
occurs if normal sinus rhythm is not quickly restored \cite{Jalife2000}.
 Separately, the sub-cellular, cellular and tissue-level electrophysiological
arrhythmia mechanisms underlying heart failure are well studied \cite{Tomaselli2004}:
common findings include altered calcium handling \cite{Bers2006}, prolongation of
the action potential duration \cite{Glukov2010}, hypertrophy \cite{Riehle2011},
gap junction remodelling \cite{Yoshida2011} and fibrosis \cite{Polyakova2011}.
However, it is less well understood how these different types of remodelling
interact at the tissue and organ levels to induce arrhythmias.

 The study of cardiac arrhythmias has been a major focus of computational biology,
as a detailed quantitative description of the underlying electrophysiology has
been developed that allows the simulation of both normal and pathological excitation
and propagation of this excitation \cite{Noble2002, Clayton2011}. As experimental
studies of cardiac arrhythmias at the tissue and organ levels have been largely
limited to voltage recordings on or near the surface of a preparation
\cite{Efimov2004,Walton2010}, computational models offer an additional research
tool that can enhance results from experimental studies \cite{Benson2011} or can
give insights into the functioning of the heart that cannot be gained through
experimental work, either due to technical limitations or ethical issues
\cite{TenTusscher2007}. The results of such simulations can be dissected in time
and space, and investigated over a broad range of parameter values, allowing a
detailed study at the cell, tissue and organ levels of the mechanisms underlying
the onset of arrhythmias \cite{Fenton1998}, and interventions aimed at either
preventing this onset \cite{Rodriguez2010,Benson2011a} or restoring normal sinus
rhythm \cite{Trayanova2006,Aslanidi2009}.

 Recently, these electrophysiology models have been coupled to mechanical models
in deforming domains in order to investigate the additional influence that
mechanics has on arrhythmogenesis \cite{Nash2004,Keldermann2010,Pathmanathan2009}.
For example, \cite{Nash2004} showed that breakdown of a re-entrant wave into a
chaotic state ({\em i.e.}\ the transition from ventricular tachycardia to
fibrillation) occurred later in a contracting compared to a non-contracting
domain. However, these studies often use either phenomenological models that,
due to their simplicity, limit the application of the model to a disease state
({\em e.g.}\ \cite{Nash2004}), or use biophysically-detailed (and therefore
computationally-demanding) models that make longer simulations intractable
without access to high-performance computing facilities ({\em e.g.}\
\cite{Keldermann2010}). An efficient, biophysically-detailed, electromechanical
model is therefore a useful research tool.

 The aims of this study were: (i) to develop a biophysically-detailed,
coupled electromechanical model of human tissue that could be used
to study re-entrant arrhythmias in both healthy and failing states; (ii) to
tailor this model to reproduce the cell- and tissue-level remodelling observed
with heart failure; and (iii) use the modified model to examine the roles of
electrophysiology (ionic currents, calcium handling, gap junctions), fibrosis
and mechanical deformation on the initiation of ventricular fibrillation in heart
failure. 

 In this paper the biophysically-detailed human electrophysiology model of ten
Tusscher and Panfilov \cite{TenTusscher2006} is coupled with a Mooney-Rivlin
mechanical model of elastic material (as in \cite{Nash2004}), leading to two
systems of partial differential equations which are approximated on unstructured,
triangular meshes in two space dimensions using the finite element method. First,
the electrophysiology and mechanics models are outlined, followed by a description
of the modifications required to distinguish failing from healthy tissue. Next,
the computational model used to approximate this system is described briefly and
its behaviour illustrated by applying it to healthy/control versions of the
electrophysiology and mechanics systems.
 This is then used for the subsequent investigation of the roles of remodelled
electrophysiology, fibrotic tissue and mechanical deformation on spiral wave
dynamics in human ventricular tissue, a key indicator in many dangerous cardiac
arrhythmias \cite{Pertsov1993}.

\section*{Methods}

 To simulate coupled cardiac electromechanical activity it is necessary to model
the propagation of the electrical waves around the heart muscle, the muscle
contraction instigated by this electrical activity and the feedback that the
electrical activity and muscle deformation have upon each other. This section will
provide an outline of the mathematical models used to represent the coupled
electromechanical activity, the computational algorithms which were used to
approximate this mathematical system, and the meshes on which the computations
were carried out.

\subsection*{Electrophysiology Model}

 For this study, a monodomain model was used to represent cardiac electrophysiology,
as described by \cite{Keener1998}
\begin{equation}
 \label{ElecPDE}
 \frac{\partial V}{\partial t} = - (I_{\mathrm{ion}} + I_{\mathrm{stim}}) + \nabla \cdot (\mathbf{D} \nabla V) ,
\end{equation}
where $\mathbf{D}$ is the diffusion tensor, $V$ is the transmembrane potential,
$I_{\mathrm{ion}}$ is the sum of all ionic currents, and $I_{\mathrm{stim}}$ is
the externally applied transmembrane current. On the boundary of the domain, zero
flux conditions are imposed, {\em i.e.}\ $\nabla V \cdot n = 0$, where $n$ is the
outward pointing normal to the boundary.
 The detailed description of the ionic currents $I_{\mathrm{ion}}$ was provided
by the ten Tusscher and Panfilov 2006 model (referred to hereafter as TP06)
\cite{TenTusscher2006}, which presents a detailed description of individual ionic
currents and intracellular ion concentrations in human epicardial ventricular cells.
This is a biophysically-detailed model, based on human electrophysiological data,
and provides the potential to simulate the conditions of end-stage tissue failure
examined later in this paper.

\subsection*{Mechanics Model}

 The cardiac tissue was modelled as a nonlinearly elastic material. The governing
stress equilibrium equations \cite{Nash2004} take the general form
\begin{equation}
 \label{stresseq}
 \frac{\partial}{\partial X_M} \left( T^{MN} F^j_N \right) = 0 ,
\end{equation}
where $T^{MN}$ is the second Piola-Kirchhoff stress tensor and $\mathbf{F}$ is the
deformation gradient tensor,
\begin{equation}
 F^j_N = \frac{\partial x_j}{\partial X_N} .
\end{equation}
 The unknown variables $x_j$ represent the deformed coordinate system, while $X_N$
are the reference coordinates, relative to which the deformed coordinates are
defined. The indices $M,\,N,\,j$ represent the coordinate axes and take the values
$1,\ldots,d$, where $d$ is the number of space dimensions. These equations are
derived from conservation of linear momentum, following Newton's laws of motion
\cite{Malvern1969}, and their solution provides the material deformation. In common
with other authors \cite{Hunter1998,Nash2000} the tissue is assumed to be
incompressible, a constraint which is enforced by imposing
\begin{equation}
 \label{eq:incompressible}
 det(\mathbf{F}) = 1 ,
\end{equation}
as in \cite{Pathmanathan2009}. On the boundary of the domain a zero normal stress is
imposed.

 In common with \cite{Pathmanathan2009}, the second Piola-Kirchhoff tensor is
split into an elastic component and a biochemical component, giving
\begin{equation}
 \label{Piola2}
 T^{MN} = \underbrace{\frac{1}{2} \left( \frac{\partial W}{\partial E_{MN}} + \frac{\partial W}{\partial E_{NM}} \right)
             - p C^{-1}_{MN}}_{\mathrm{elastic}} + \underbrace{T_a C^{-1}_{MN}}_{\mathrm{biochemical}} ,
\end{equation}
where $W$ is the scalar strain energy function, $\mathbf{E} = \frac{1}{2} ( \mathbf{C} - \mathbf{I} )$
is the Lagrange-Green strain tensor ($\mathbf{C} = \mathbf{F}^T \mathbf{F}$ being
Green's strain tensor), $p$ is a Lagrange multiplier that is used to enforce the
incompressibility constraint (\ref{eq:incompressible}) and is an additional unknown
variable commonly referred to as the pressure, and $T_a$ is the active tension
generated by the electrophysiology. Following \cite{Nash2004}, the strain energy
function for Mooney-Rivlin materials \cite{Mooney1940,Rivlin1948} is used. This is
given by
\begin{equation}
 \label{MooneyRivlin}
 W(I_1,I_2) = c_1 ( I_1 - 3 ) + c_2 ( I_2 - 3 ) ,
\end{equation}
where $I_1 = tr \, \mathbf{C}$ and $I_2 = \frac{1}{2} [ ( tr \, {\bf C} )^2 - tr \, \mathbf{C}^2 ]$
are the first and second invariants of the Cauchy-Green deformation tensor ($tr$
being the trace of the tensor $\mathbf{C}$, the sum of its diagonal elements).
Throughout this work, the parameters $c_1$ and $c_2$ were set to $2\,\mathrm{kPa}$
and $6\,\mathrm{kPa}$, respectively.
 It has been shown experimentally that cardiac tissue exhibits different responses
along the various material axes \cite{Smaill1991} so, since the chosen strain
energy function is isotropic, the biochemical component in Equation (\ref{Piola2})
is set so that the active tension acts in only one direction, coinciding with the
major axis of the diffusion tensor $\mathbf{D}$, as described in \cite{Pathmanathan2009}.
This ensures that the forces generated by the biochemical activity act along the
direction of fibre orientation \cite{Nash2000,Whiteley2007}. 

 The last component of Equation (\ref{Piola2}) governs the response of the mechanics
to the active tension ($T_a$) generated by the biochemical activity in the tissue.
In \cite{Nash2004} (Equations (22c) and (23), subsequently updated at {\tt www.cellml.org}),
a phenomenological description of the tension within cardiac tissue is provided,
in which the active tension is determined by the transmembrane voltage, {\em i.e.}\
we use
\begin{equation}
 \label{eq:ActiveTension}
 \frac{d T_a}{d t} = \epsilon(V_t) ( K_{T_a} V_t - T_a ) ,
\end{equation}
where $K_{T_a} = 9.58\,\mathrm{kPa}$ is a parameter which controls the maximum
amplitude of the active tension ($T_a$), $V_t$ is the normalised transmembrane
voltage (scaled linearly so that $0 \leq V_t \leq 1$), and
\begin{equation}
 \label{eq:ActiveTension2}
 \epsilon(V_t) = \left\{ \begin{array}{ll}
                         10 \epsilon_0 & \mathrm{for} \;\; V_t  <   0.005 \\
                            \epsilon_0 & \mathrm{for} \;\; V_t \geq 0.005 .
                        \end{array} \right.
\end{equation}
The value of $\epsilon_0$ was taken to be 1.
 Note that in \cite{Kirk2012} (see Section 7.3) the active tension was determined
by the calcium transient instead of the transmembrane voltage. This has a much
steeper wave descent which slightly reduces the overall tissue deformation. However,
it did not change the conclusions that could be drawn about the effects of tissue
remodelling on spiral wave stability.

\subsection*{Simulating Heart Failure}
\label{sec:ModelHeartFailure}

 Electrophysiological arrhythmia mechanisms underlying cardiac disease are well
studied \cite{Tomaselli2004}, with common findings including altered calcium
handling \cite{Bers2006} and prolongation of the action potential duration
\cite{Glukov2010}. To simulate failing tissue, some components of the ionic
currents, $I_{\mathrm{ion}}$ in Equation (\ref{ElecPDE}), were amended from those
presented in the original ten Tusscher and Panfilov model \cite{TenTusscher2006}:
the transient outward current ($I_{\mathrm{to}}$) maximal conductance was reduced
by 48\% \cite{Kaab1998}, the inward rectifier potassium current ($I_{\mathrm{K1}}$)
was reduced by 44\% \cite{Beuckelmann1993}, the sodium-potassium pump current
($I_{\mathrm{NaK}}$) was reduced by 40\% \cite{Schwinger2003}, the sodium-calcium
exchanger current ($I_{\mathrm{NaCa}}$) was increased by 80\% \cite{Hasenfuss1999}
and sarcoplasmic reticulum uptake current ($I_{\mathrm{up}}$) reduced by 30\%
\cite{Jiang2002,Dash2001}.
 These changes, referred to from now on as the remodelled electrophysiology, are
designed to simulate failing tissue by increasing the resting membrane potential
and prolonging duration.

 In heart failure, gap junctions (that allow current flow between cells) are
reorganised, and instead of being principally located at the ends of the cells,
they become ``lateralised" so that there is an increase in transverse gap junction
numbers and a decrease in longitudinal gap junction numbers within a single cell.
However, because of the changes in cell size with the hypertrophy that accompanies
heart failure, there is no net change in gap junction numbers in the transverse
direction, but still a net decrease (of between 28\% \cite{Xun2005} and 40\%
\cite{Dupont2001}) in longitudinal gap junction numbers \cite{Jongsma2000}.
 Gap junction changes will affect cell resistivity but, because the electrophysiology
model, Equation (\ref{ElecPDE}), treats the tissue as a continuum, this resistivity
is made up of a ``myoplasmic resistivity'' (caused by the structures inside the
cell) and a ``junctional resistivity'' (caused by the gap junctions between cells).
We assume that junctional resistivity accounts for 22\% of total resistivity
\cite{Thomas2003}  and that a reduced gap junction expression (33\% for our
simulations) is followed by a corresponding increase in junctional resistivity
\cite{Thomas2003}.
 Thus, the cumulative effects of gap junction remodelling is introduced simply by
modifying the diffusion tensor in Equation (\ref{ElecPDE}). Diffusion in control
tissue was set to $0.154\,\mathrm{cm}^2/\mathrm{ms}$ along the fibre axis, and
reduced nine-fold (to reduce conduction velocity three-fold) in the cross-fibre
direction. In heart failure, diffusion in the fibre direction was reduced to
$0.139\,\mathrm{cm}^2/\mathrm{ms}$, with cross-fibre diffusion remaining at its
control level.

 Fibrotic tissue within the cardiac muscle contributes to an increase in the
incidence of atrial and ventricular arrhythmias \cite{Everett2007,Assomull2006}.
Exactly how the fibrosis contributes to the generation of arrhythmias is unknown,
but impaired electrical conduction is a significant contributory factor
\cite{TenTusscher2007a,Spach2007}. In the normal healthy heart, approximately 6\%
of cardiac muscle is made from extra-cellular connective tissue
\cite{Rossi2001,TenTusscher2007a}. However, in a diseased heart, there is increased
formation of fibrotic cells which increases the percentage of connective tissue to
between 10\% and 35\% \cite{Rossi2001,TenTusscher2007a}. To simulate the effects
of diffuse fibrosis a large number of small areas within the domain were set to
be inexcitable, as in \cite{TenTusscher2007a}.
 In these simulations approximately 27\% of the tissue was designated inexcitable,
within the range seen experimentally \cite{Rossi2001,TenTusscher2007a}. These areas
remain in the mechanics system, but do not contract due to the electrical wave
since they are inexcitable. Hence they have no electrically-activated tension
and do not contribute to the active tissue deformation, instead maintaining their
passive deformation properties.

\subsection*{Computational Methods}

 This section provides a brief outline of the computational algorithms chosen and
further details of their implementation for this particular application can be
found in \cite{Kirk2012}. Both the electrophysiology and the mechanics were
approximated on unstructured triangular meshes using the finite element method
(FEM). Although the computational domains used in this work were all rectangular,
the use of unstructured triangles provides flexibility for future simulations on
the complex geometries required for realistic representation of a whole ventricle
or a complete heart. Furthermore, the use of unstructured meshes avoids the
systematic introduction of mesh-induced anisotropy, as illustrated by the mild
squaring of the wave-fronts visible in the numerical results presented in
\cite{TenTusscher2006}, a phenomenon which is exaggerated by the steepness of the
wave-fronts generated by the TP06 model.

 The time-dependent equation governing the electrophysiology, Equation (\ref{ElecPDE}),
was discretised using the Galerkin FEM. The spatial variation of the unknown
variable ($V$) was represented by a continuous, piecewise linear, function
(providing second order accuracy in space) where the unknowns of the discrete
system are the values of this function at the nodes of the triangular mesh
\cite{Gresho2000}. The temporal variation of these discrete nodal values was
approximated using a semi-implicit approach in which the Crank-Nicolson method
was applied to the diffusion term (for which it is unconditionally stable, so
the time-step can be chosen independently of the prohibitive stability restrictions
imposed on explicit schemes when the spatial mesh is refined, and second order
accurate \cite{Thomas1995}) and the reaction term was approximated using an explicit
forward Euler method. The sparse system of linear equations resulting from the
implicit FEM approximation was solved using an ILU-preconditioned GMRES solver
\cite{Saad1986}, as implemented in the SPARSKIT software
({\tt www-users.cs.umn.edu/\verb=~=saad/software/SPARSKIT/}). The forward Euler
method was also applied to the evolution of the active tension, Equation
(\ref{eq:ActiveTension}).

 After a specified number of time-steps (10 in this work) of the electrophysiology
system, the updated active tension was used to produce a new estimate of the
deformation of the tissue via the stress equilibrium equations, (\ref{stresseq})
coupled with (\ref{eq:incompressible}). These were also approximated using the
FEM on an unstructured triangular mesh, though the nature of this mathematical
system necessitated a slightly different approach. The unknown deformed coordinate
system (defined by $x_j, \; j=1,2$) was represented by a pair of continuous,
piecewise quadratic, functions where the unknowns of the discrete system are the
values of these functions at both the nodes and the edge-midpoints of the triangular
mesh, but the unknown pressure variable ($p$) was, like $V$, approximated by a
continuous, piecewise linear, function with the unknown values at the mesh nodes.
This additional level of complexity in the representation is required to produce
a stable algorithm \cite{Gresho2000}. The resulting, highly nonlinear, discrete
system was solved with a matrix-free Newton-Krylov iterative method \cite{Kelley1987},
in which the GMRES algorithm \cite{Saad1986} was used to solve the linear system
which occurs at each iteration of the Newton method for solving the nonlinear system.
The initial estimate required by the Newton method was provided by linear
extrapolation from the two previously calculated sets of discrete deformed
coordinates.
 The efficiency of the mechanics solver was significantly enhanced by applying
an ILUT preconditioner \cite{Saad1996,Chen2005}. This typically improved
performance by a factor of more than 20 \cite{Kirk2012}.
 We note though that, while this is a significant improvement, the approach
is not optimal, and that in the future more sophisticated preconditioners, based
on the multigrid approach, have the potential to provide iterative convergence
rates that are independent of the system size.
 To prevent spurious rotation and translation, and hence preserve uniqueness of
the numerical approximation of the tissue deformation, the mesh node closest to
the centre of the domain was fixed in space and a neighbouring node selected to
be at the same vertical height, was not allowed to move in the vertical direction
\cite{Kirk2012}.

 In order to resolve the steep wave fronts which occur within the electrical waves
produced by the TP06 model and, consequently, to produce a numerically-converged
approximation of the wave speed, it was necessary to approximate the electrophysiology
system on a fine mesh \cite{TenTusscher2006,Kirk2012}. In contrast, the variations
in the tissue deformation are much smoother, so the mechanics system does not need
to be approximated on such a fine mesh to produce numerically converged results
\cite{Pathmanathan2009,Kirk2012}. This is fortuitous because it is the solution of
the discrete mechanics system which is the bottleneck in the computation: if the
same computational mesh is used it typically takes 100s-1000s times longer to
solve the mechanics system than the electrophysiology system, even with
preconditioning enabled \cite{Kirk2012}. 
 Therefore, the electrophysiology and mechanics systems were solved on two
different meshes and, in order to simplify the communication between the two
numerical solvers, the electrophysiology mesh was derived by repeated uniform
refinement of the mechanics mesh, {\em i.e.}\ one mesh is embedded within the
other so no interpolation is required to couple the two components, although the
active tension ($T_a$) must be integrated over multiple elements of the
electrophysiology mesh in order to evaluate it on the mechanics mesh when
approximating Equations (\ref{stresseq}) and (\ref{Piola2}).

 The electrophysiology and gap junction remodelling were both introduced into the
computational model by changing the appropriate physiological parameters. The
fibrotic tissue manifests itself as many small regions of non-conduction within
the electrophysiology model, which were represented by removing patches of
computational elements from the electrophysiology meshes, and applying Neumann
no-flux conditions at the boundaries of these patches \cite{TenTusscher2007a}.
The positions of the patches were chosen randomly, though care was taken to ensure
that they were separated by at least one layer of mesh elements. No modification
to the mechanics mesh was necessary since these patches simply provide zero
contribution to the active tension which determines the local tissue deformation.

 The combination of using (i) a coarse mechanics mesh embedded in a finer
electrophysiology mesh, (ii) implicit time-stepping for the diffusion terms,
(iii) multiple time-steps of the electrophysiology between updates of the mechanical
deformation, and (iv) ILU preconditioning for both electrophysiology and mechanics
systems, massively reduced the CPU time required to undertake these simulations.
 With this approach,
our solver is
efficient enough that the simulations of re-entry presented in this paper remained
tractable on a desktop PC (Intel Xeon E5420, 2.5 gigahertz clock speed, 4 gigabytes
memory).

\section*{Results}

 The results presented in this section are chosen to illustrate the effects on the
stability of two-dimensional spiral waves, on both static and deforming domains,
of: (i) remodelled electrophysiology; (ii) gap junction remodelling and fibrosis;
and (iii) a combination of both.

\subsection*{Test Cases and Initialisation}
\label{sec:init}

 The simulations were performed on unstructured triangular meshes covering a
$120\,\mathrm{mm}\,\times\,120\,\mathrm{mm}$ square domain. In all cases an
initial, stable, spiral wave was generated by first applying a $52\,\mathrm{mV}$
stimulus for $1\,\mathrm{ms}$ to the left face of the domain (using
$I_{\mathrm{stim}}$ in Equation (\ref{ElecPDE})) to initiate a planar wave-front
propagating perpendicular to that face. As this wave traverses the middle of the
domain (at $115\,\mathrm{ms}$), the voltage in the whole of the lower half of the
domain was overwritten and fixed to the resting potential for a short period of
time ($35\,\mathrm{ms}$ in this case), after which a spiral wave evolves around
the end of the wave-front which has been created in the centre of the domain. The
simulation was continued for $5000\,\mathrm{ms}$ to ensure that the spiral wave
dynamics had settled down to a regular periodic behaviour. This was carried out
with the parameters of the TP06 model of electrophysiology in healthy tissue
\cite{TenTusscher2006} chosen to simulate a restitution slope of 1.1, providing
stable spiral wave structures with which to initialise all of the simulations
presented in this paper. Separate initialising computations were needed for each
different computational mesh used and computations which included electromechanical
coupling were initialised on deforming domains. This allowed investigation of whether
or not the modified conditions (deformation, remodelling, fibrosis) would destabilise
an initially stable spiral wave structure. For the simulations involving fibrosis
the regions of non-conducting tissue, introduced by removing patches of elements
from the computational mesh, were present during the initialisation as well as the
subsequent simulation.

\subsection*{Electrophysiology}
\label{sec:elec}

 In order to provide a set of control results with which later simulations can be
compared, simulations on a static domain were carried out with electrophysiology
represented by the TP06 model \cite{TenTusscher2006} and an anisotropic diffusion
tensor providing a diffusion rate of $0.154\,\mathrm{cm}^2/\mathrm{ms}$ along the
fibre axis, acting diagonally from bottom-left to top-right across the domain, and
reduced nine-fold in the cross-fibre direction.
 The unstructured triangular mesh used (made up of 634368 triangles and 318065
nodes) had an approximate average edge-length of $0.21\,\mathrm{mm}$. This is
comparable to the regular finite difference mesh used in \cite{TenTusscher2004},
in which the nodes were a distance $0.2\,\mathrm{mm}$ apart. The time-stepping
for the approximation to Equation (\ref{ElecPDE}) was carried out with a fixed
time-step of $0.08\,\mathrm{ms}$. The mesh on which the tissue deformation was
approximated in the following computations was considerably coarser (with only
2478 triangles, equating to 11429 degrees of freedom).

\begin{figure}
 \begin{center}
  \mbox{{\bf A} \hspace{4.8cm} {\bf B} \hspace{4.8cm} {\bf C} \hspace{4.8cm}} \\ \vspace{-\baselineskip}
  \label{fig:APDNorm11}\includegraphics[width=5.2cm]{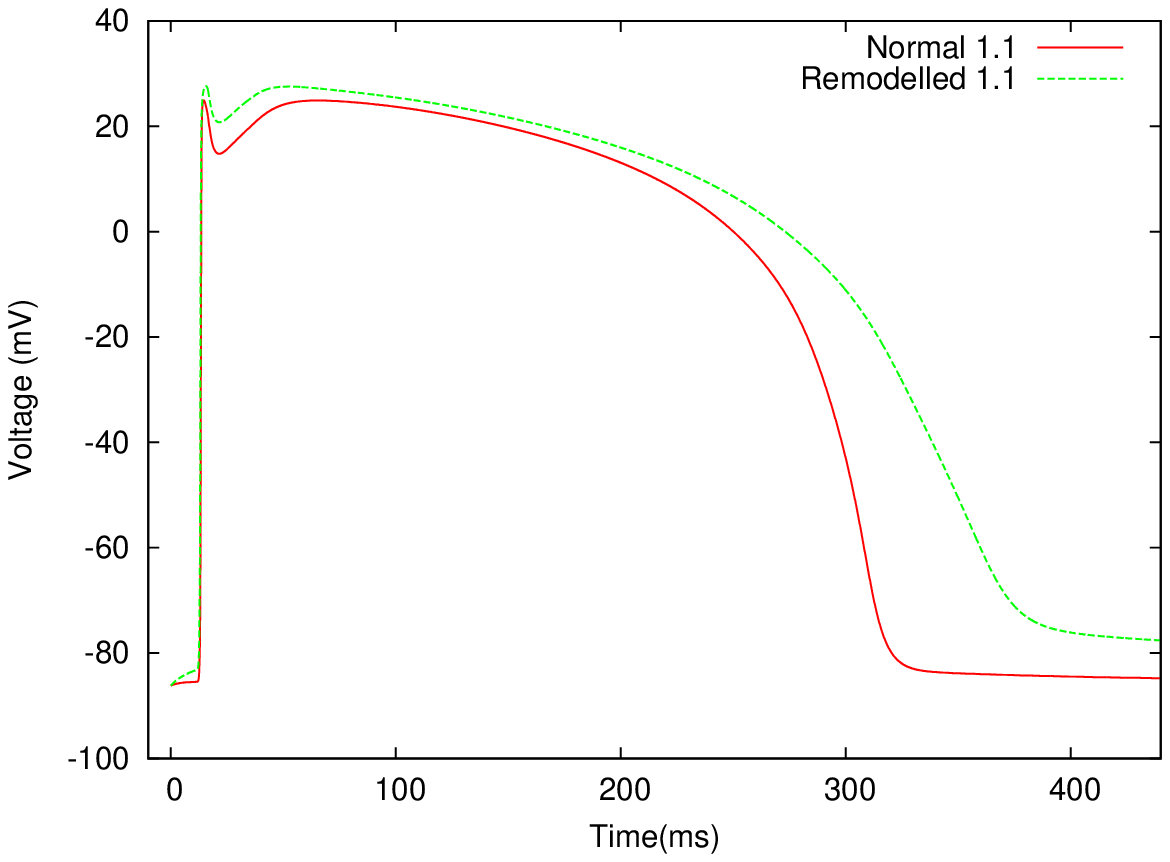}
  \label{fig:APDNorm14}\includegraphics[width=5.2cm]{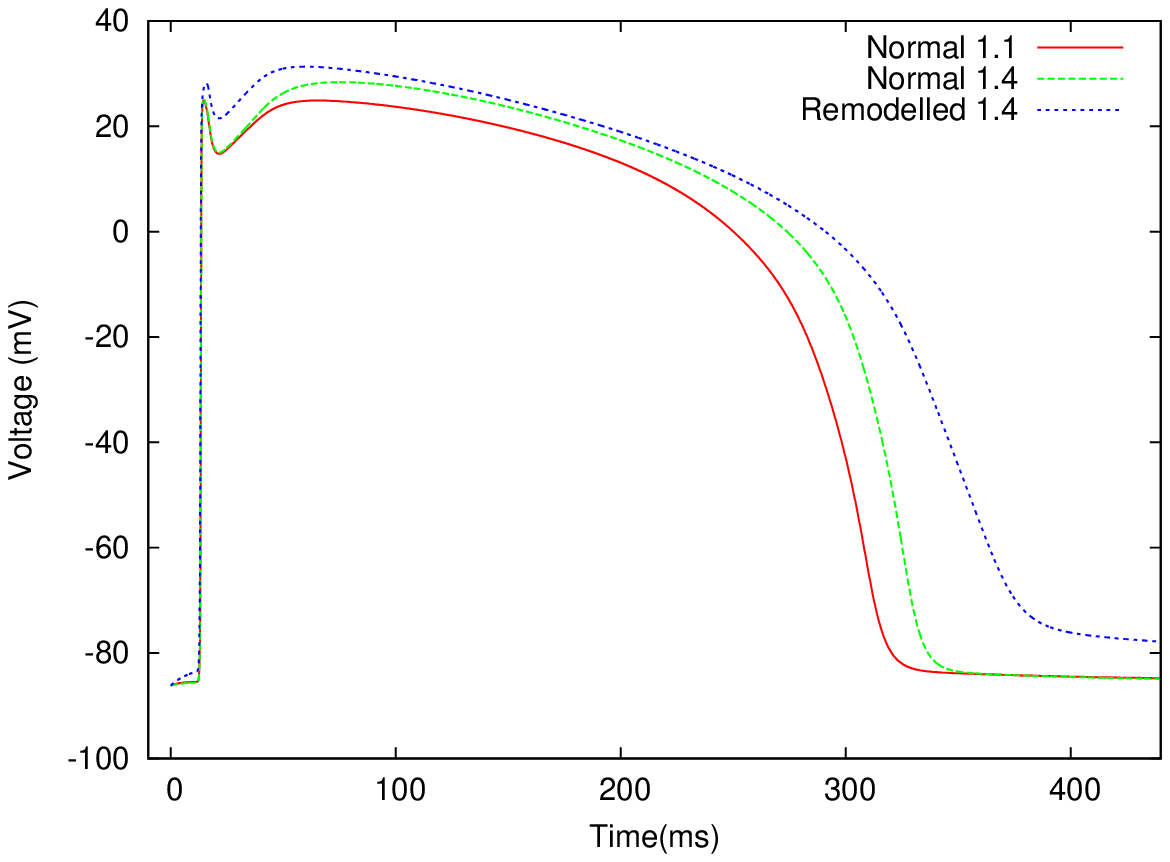}
  \label{fig:APDNorm18}\includegraphics[width=5.2cm]{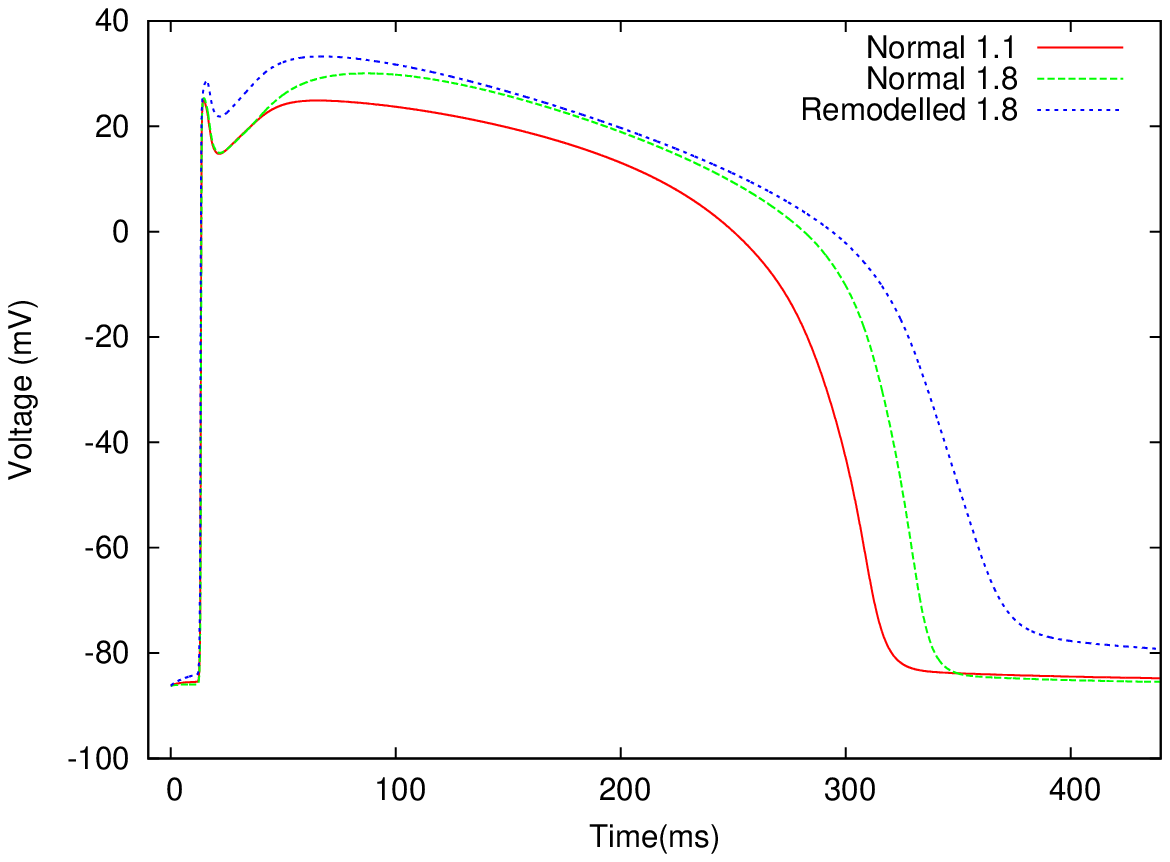}
  \caption{{\bf Effect of restitution slope on transmembrane voltage.} Comparison
   of transmembrane voltage profiles for the TP06 model, sampled at a single node
   of the unstructured computational mesh, for normal and remodelled electrophysiology:
   ({\bf A}) restitution slope of 1.1; ({\bf B}) restitution slope of 1.4; ({\bf C})
   restitution slope of 1.8.}
  \label{fig:Rest1}
 \end{center}
\end{figure}

\begin{figure}
 \begin{center}
  \mbox{{\bf A} \hspace{6.0cm} {\bf B} \hspace{8.3cm}} \\ \vspace{-\baselineskip}
  \includegraphics[height=6.0cm]{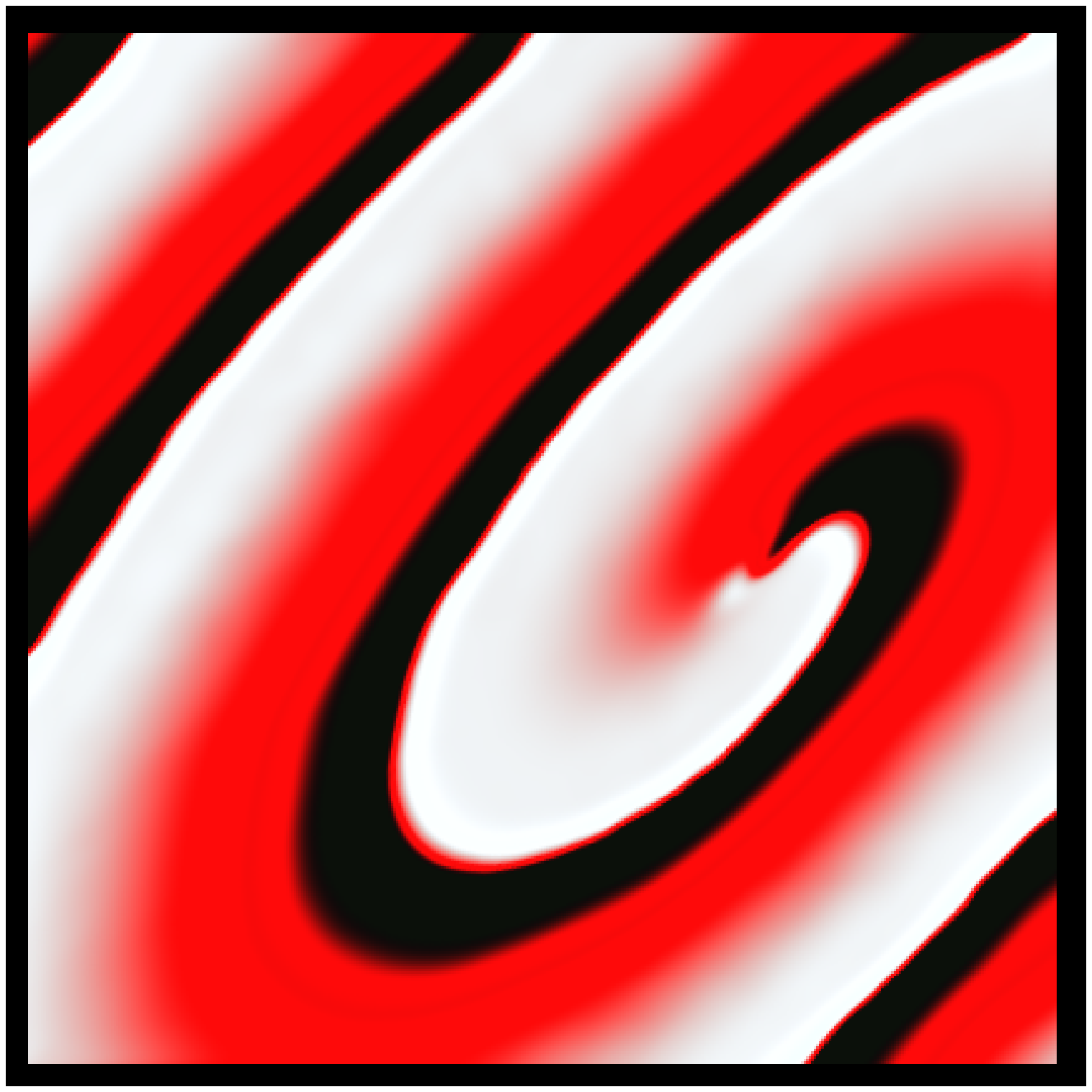} \quad
  \includegraphics[height=6.0cm]{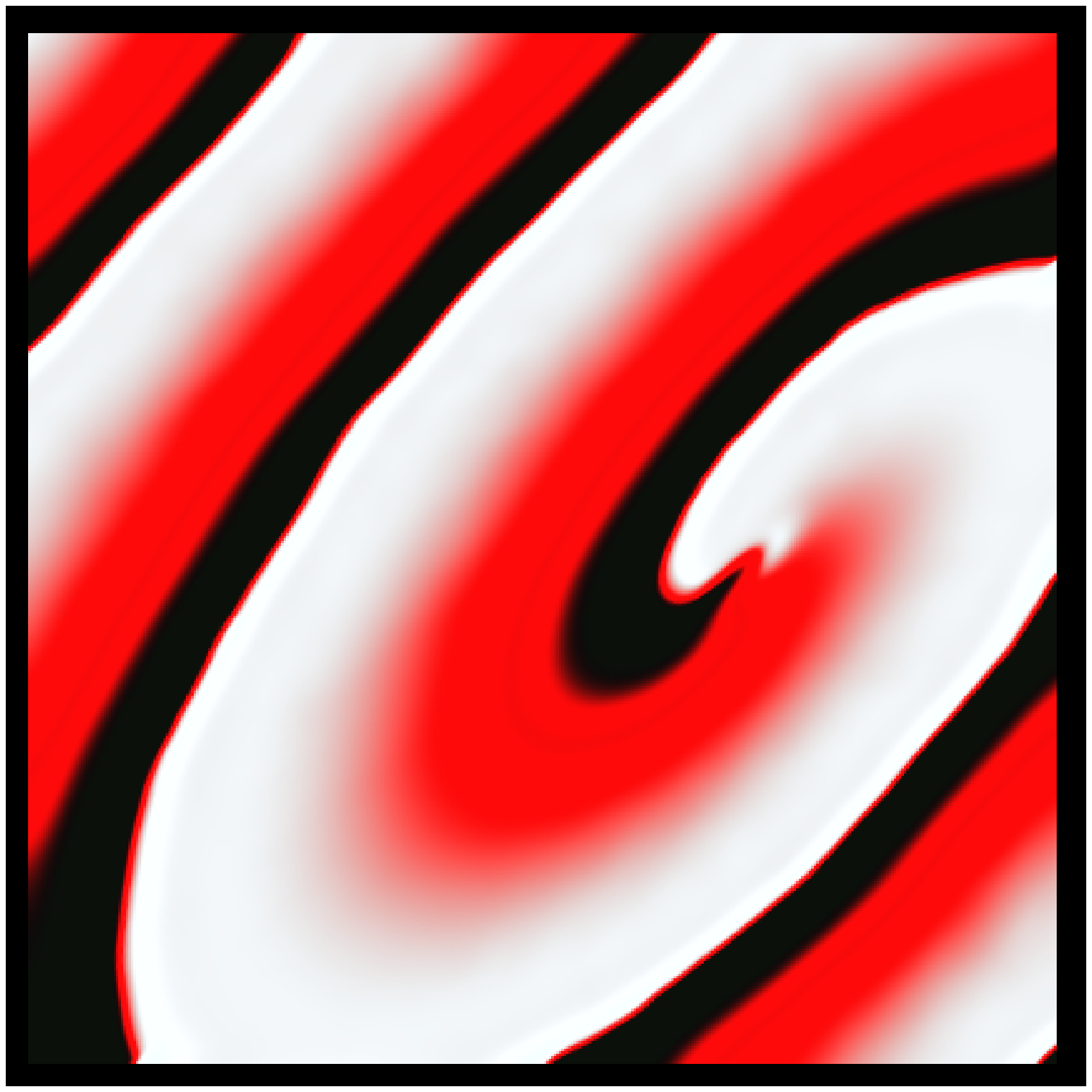} \quad 
  \includegraphics[height=6.0cm, width=1.5cm]{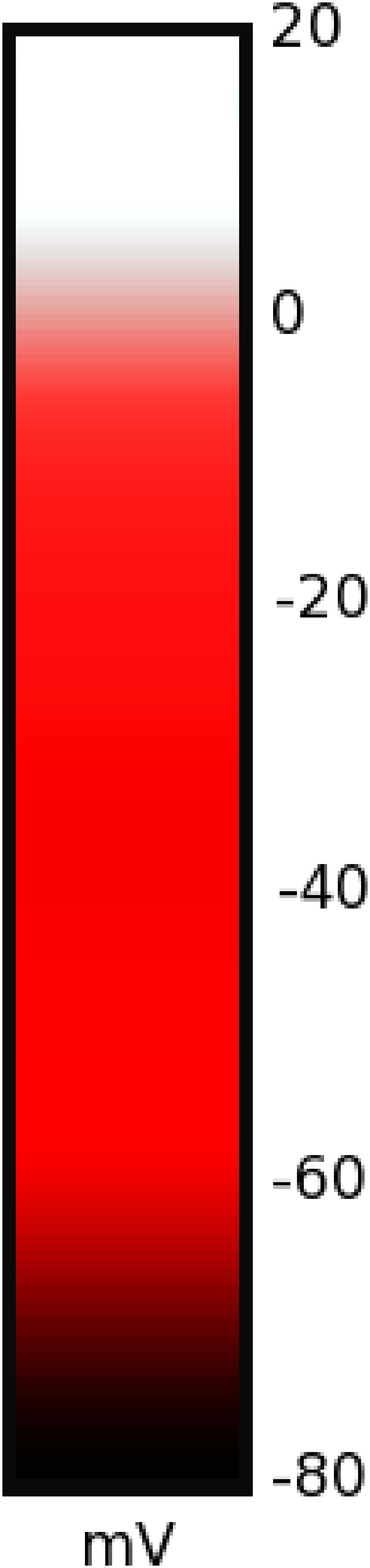} \\
  \vspace{4mm}
  \mbox{{\bf C} \hspace{6.0cm} {\bf D} \hspace{8.3cm}} \\ \vspace{-\baselineskip}
  \includegraphics[width=6.0cm]{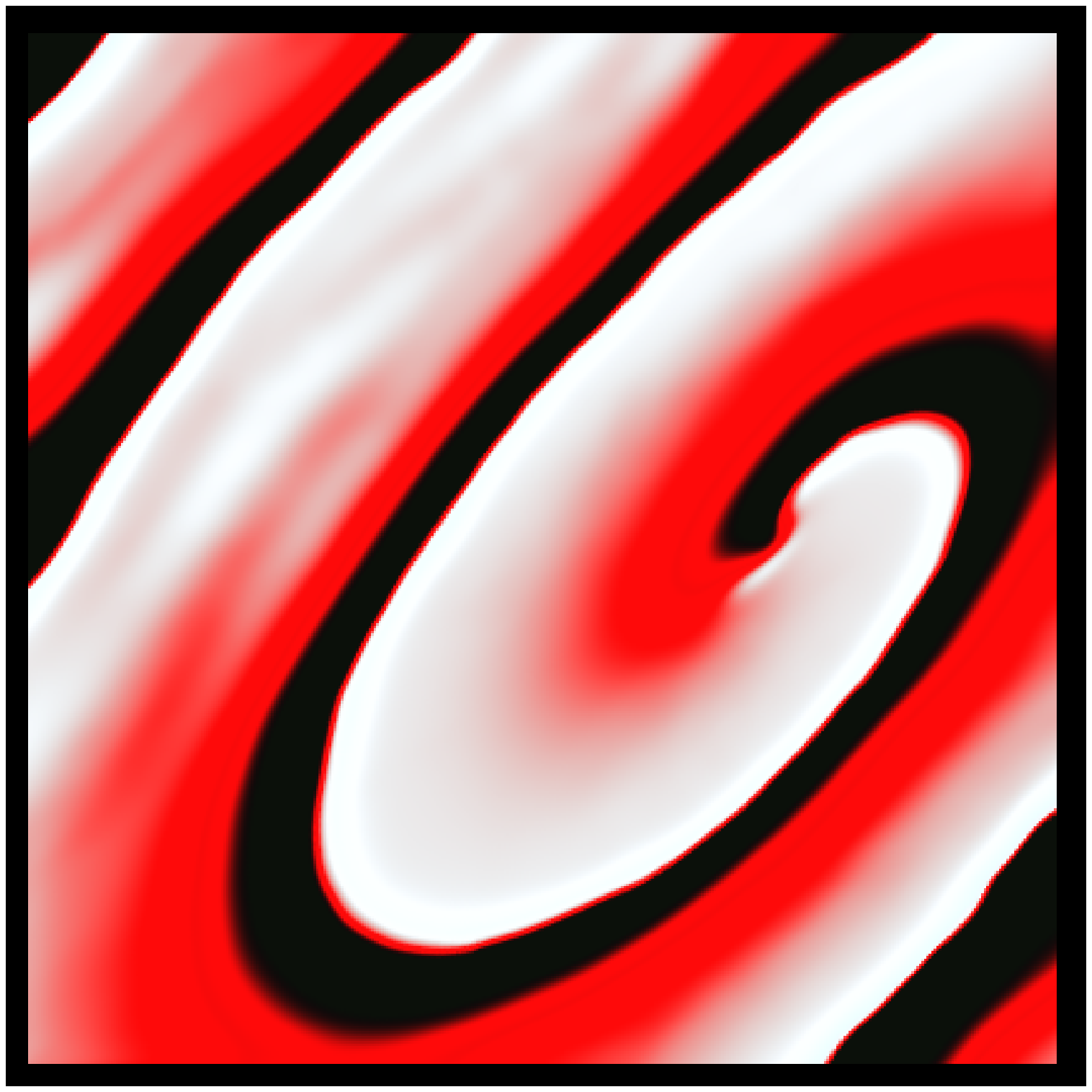} \quad
  \includegraphics[width=6.0cm]{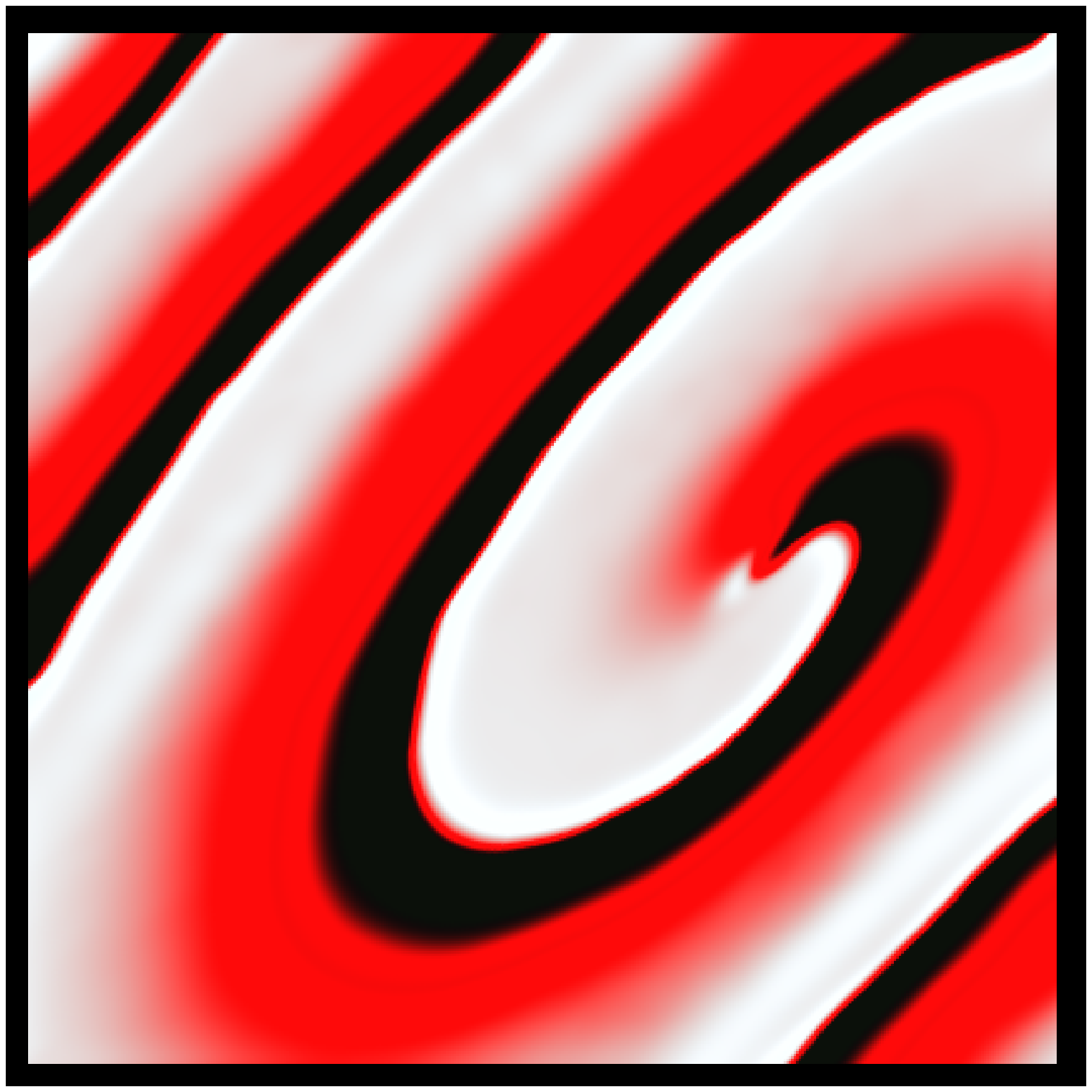} \quad
  \includegraphics[height=6.0cm, width=1.5cm]{./ColourScale6.eps} \\
  \vspace{4mm}
  \mbox{{\bf E} \hspace{6.0cm} {\bf F} \hspace{8.3cm}} \\ \vspace{-\baselineskip}
  \includegraphics[width=6.0cm]{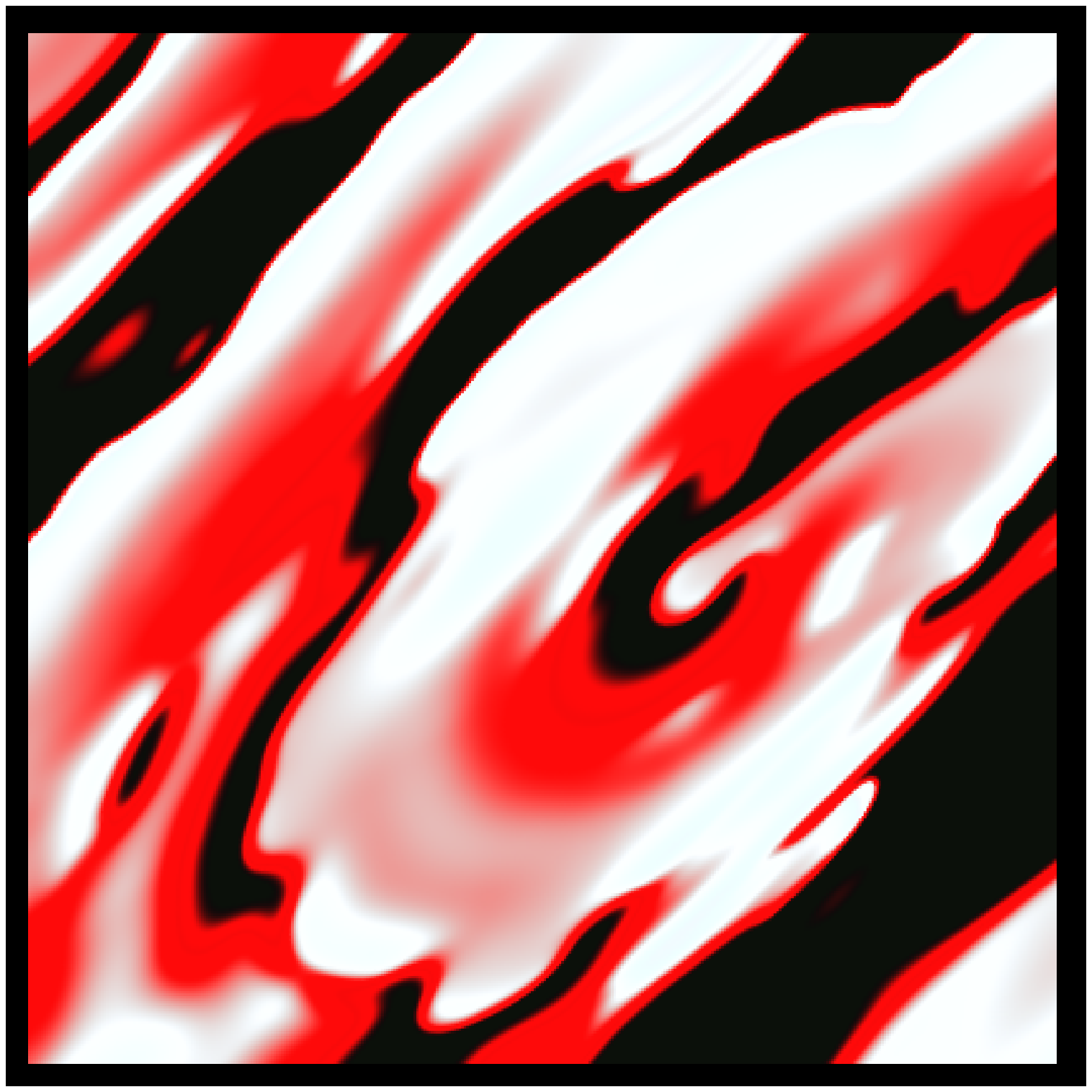} \quad
  \includegraphics[width=6.0cm]{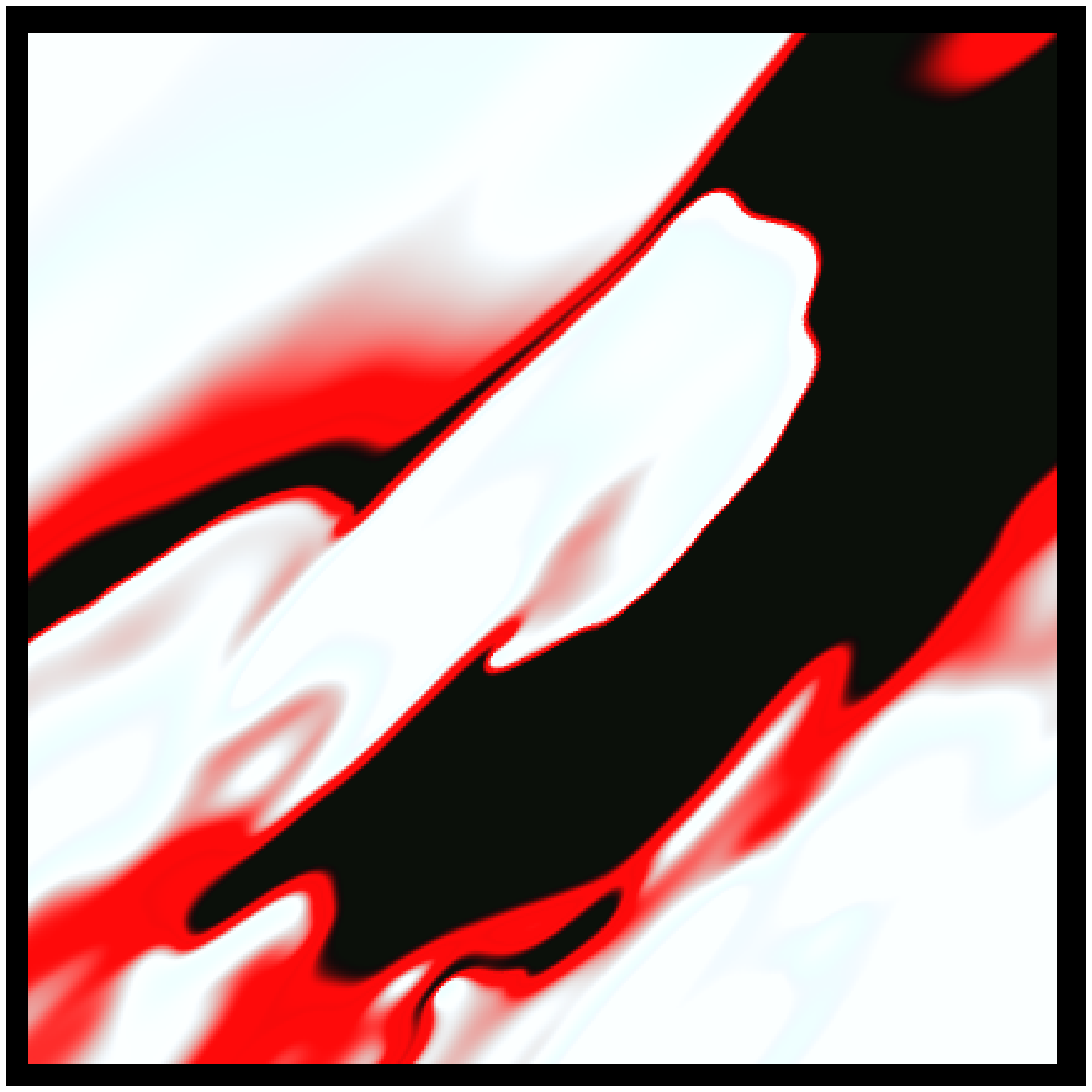} \quad
  \includegraphics[height=6.0cm, width=1.5cm]{./ColourScale6.eps} \\
  \caption{\label{fig:TP06-18} {\bf Control experiments.} Snapshots of spiral wave
   dynamics for control tissue without deformation. Each row illustrates the voltage
   at $t = 1000\,\mathrm{ms}$ (left) and $t = 3000\,\mathrm{ms}$ (right). Results
   are shown for restitution slopes of 1.1 ({\bf A}, {\bf B}), 1.4 ({\bf C},
   {\bf D}) and 1.8 ({\bf E}, {\bf F}).}
 \end{center}
\end{figure}

 Simulations were undertaken using dynamic restitution slopes of 1.1, 1.4 and 1.8
(as defined in Table 2 of \cite{TenTusscher2006}). The changes in the restitution
slope alter the action potential profile: the higher-valued slopes correspond to
increased resting potentials, higher plateaux and prolonged duration, as can be
seen by comparing the panels of Figure \ref{fig:Rest1}.
 Snapshots of the results of these simulations, shown in Figure \ref{fig:TP06-18},
demonstrate qualitatively similar behaviour to that seen in Figure 7 of
\cite{TenTusscher2006}. In particular, with their ``standard'' $I_{\mathrm{Na}}$
dynamics (the first column of that figure), the initially stable spiral wave is
maintained with restitution slopes of 1.1 and 1.4, but with a restitution slope
of 1.8 the spiral wave breaks up into a chaotic state. For a restitution slope of
1.4, panels {\bf C} and {\bf D} of Figure \ref{fig:TP06-18} show that small
perturbations to the voltage profile, due to the transition from an initial state
derived from a spiral wave produced using a restitution slope of 1.1, are still
visible after $1000\,\mathrm{ms}$, but have been smoothed out by the time the
simulation has reached $3000\,\mathrm{ms}$. The spiral wave becomes progressively
less stable as the restitution slope increases.

 The effect of coupling the electrophysiology and mechanics components of the model
was investigated in \cite{Kirk2011}, which presented the results obtained by
repeating these numerical experiments on a deforming domain. The snapshots presented
in Figure 3 of that paper are indicative of the tendency for the inclusion of
mechanical response and tissue deformation (in the case where the TP06
electrophysiology model is coupled with the Mooney-Rivlin mechanical model) to
destabilise spiral waves: for a dynamic restitution slope of 1.4 the spiral wave
which is stable in a static domain is unstable in a deforming domain.

\subsection*{Remodelled Electrophysiology}
\label{sec:remodel}

\begin{figure}
 \begin{center}
  \mbox{{\bf A} \hspace{7.1cm} {\bf B} \hspace{7.6cm}} \\ \vspace{-\baselineskip}
  \label{fig:BaseC18}\includegraphics[height=6.7cm]{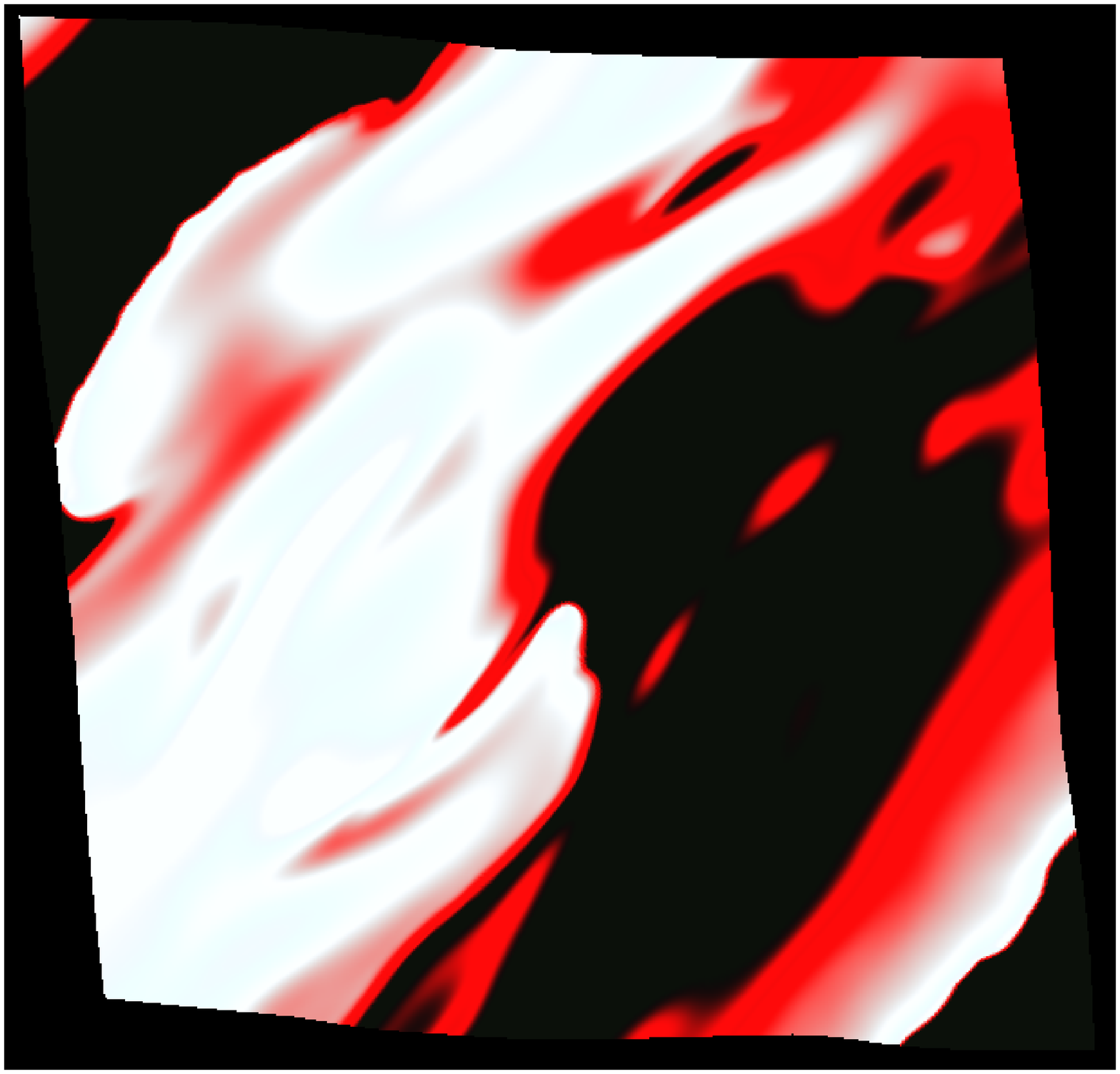} \quad
  \label{fig:CurrC18}\includegraphics[height=6.7cm]{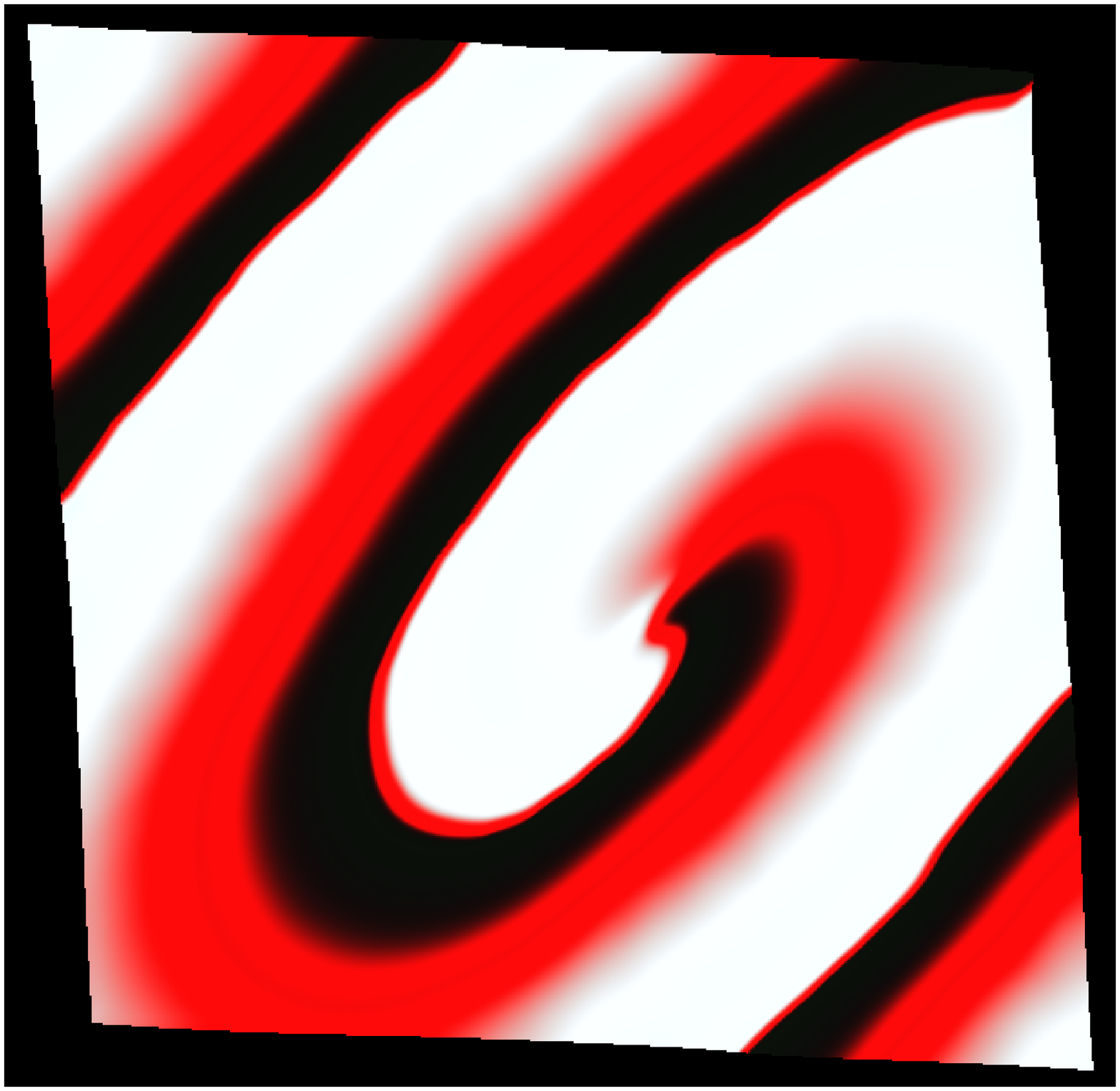}
  \caption{{\bf Effects of remodelled electrophysiology.} Snapshots of spiral wave
   dynamics for a deforming domain with a restitution slope of 1.8, taken at
   $t = 7000\,\mathrm{ms}$ with ({\bf A}) control electrophysiology and ({\bf B})
   remodelled electrophysiology.}
  \label{fig:CurrC}
 \end{center}
\end{figure}

 Figure \ref{fig:CurrC} illustrates the effect that remodelling the electrophysiology
to represent failing tissue has on spiral wave dynamics. The snapshots shown focus
on the simulation with restitution slope 1.8, since this is the only one in which
the stability of the spiral wave has been visibly altered. In this case a spiral
wave which clearly broke up in the control tissue, on both static and deforming
domains, remains stable when the remodelled electrophysiology was introduced. The
remodelling also stabilised the spiral wave when it was modelled on a static
domain (results not shown here). When the restitution slope was fixed at 1.1 or
1.4 the spiral wave was already stable in the control tissue and this stability
was not disturbed by the remodelling.

\subsection*{Gap Junction Remodelling and Fibrosis}

 The effects of changes in tissue properties, represented by the introduction of
gap junction remodelling and regions of non-conducting tissue, in the absence of
electrophysiology remodelling were investigated using a series of cases, summarised
in Tables \ref{tab:meshes1} and \ref{tab:FibAreasLower}, in which the parameters
relating to the size and distribution of the patches of inexcitable tissue,
introduced to represent diffuse fibrosis, were varied.
 Both tables show stability on static domains: all situations were unstable when
the analogous simulations were carried out on deforming domains.

\begin{table}
 \begin{center}
  \begin{tabular}{||l||c||c|c|c|c|c||} \hline
   Mesh                  &   N0   &           F0          &           F1          &           F2          &           F3          \\ \hline
   Patch Area            &  N/A   & $8.72\,\mathrm{mm}^2$ & $1.93\,\mathrm{mm}^2$ & $0.53\,\mathrm{mm}^2$ & $0.27\,\mathrm{mm}^2$ \\ \hline
   Spiral Wave Stability & Stable &        Unstable       &        Unstable       &        Unstable       &         Stable        \\ \hline
  \end{tabular}
  \caption{{\bf The computational domains used for the fibrosis simulations to
   investigate the effect of varying the individual patch size.} The first column
   contains the mesh used for simulating normal tissue, {\em i.e.}\ without fibrotic
   regions. The subsequent columns describe meshes in which the size of the individual
   fibrotic regions (patch area) is varied. In each case the number of fibrotic
   regions is chosen so that approximately 27\% of the mesh area is removed randomly
   to represent the non-conducting fibrotic areas. The final row indicates the
   stability of a spiral wave in a static domain when the restitution slope is 1.4.
   All meshes produced an unstable spiral wave in a deforming domain.}
  \label{tab:meshes1}
 \end{center}
\end{table}

 Initially, spiral wave stability was investigated on static domains, and the size
of each patch of fibrotic tissue was varied, as described in Table \ref{tab:meshes1}.
In all cases, approximately 27\% of the tissue was designated as inexcitable in total.

\begin{figure}
 \begin{center}
  \mbox{{\bf A} \hspace{6.7cm} {\bf B} \hspace{7.1cm}} \\ \vspace{-\baselineskip}
  \label{fig:BaseU14}\includegraphics[height=6.7cm]{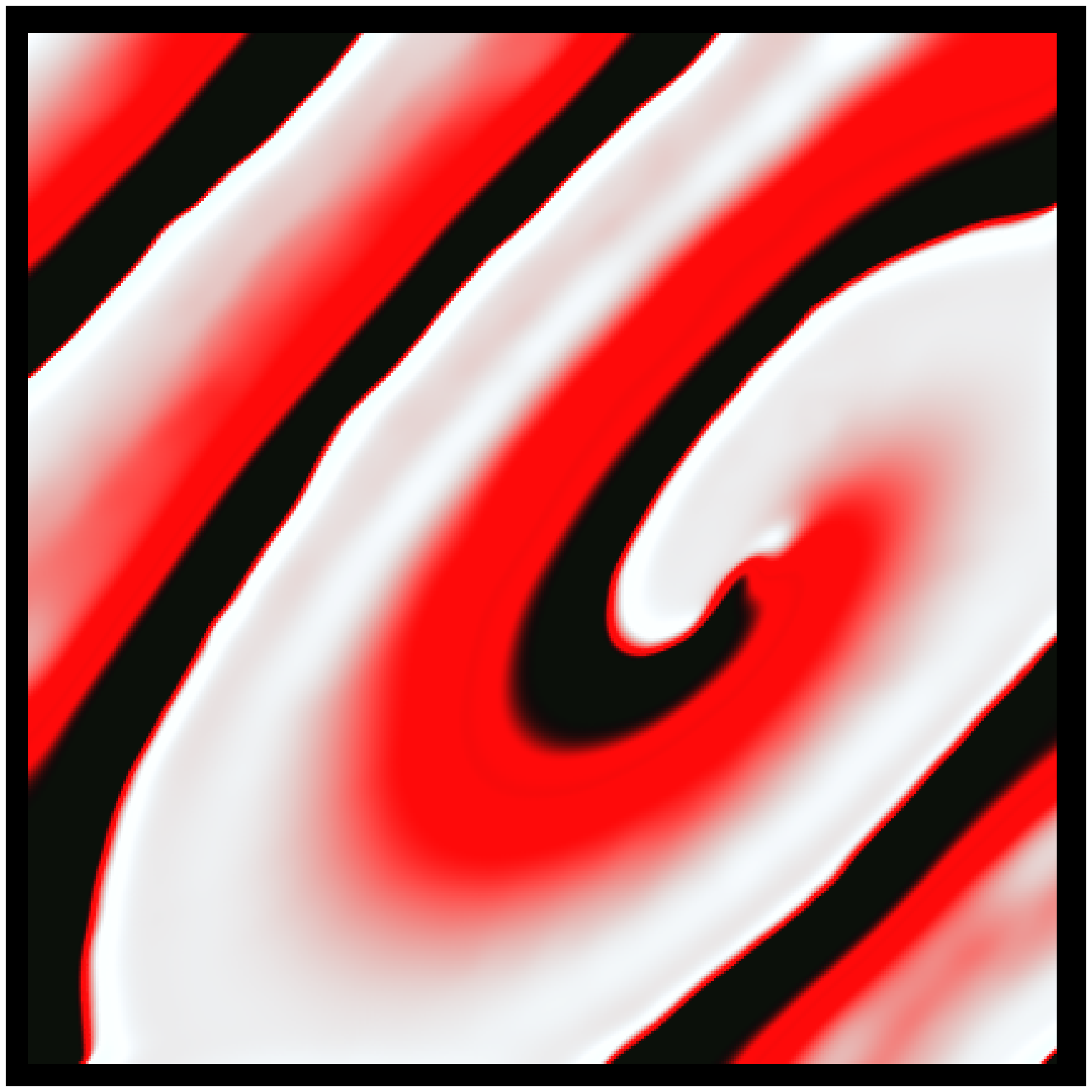} \quad
  \label{fig:FibU14}\includegraphics[height=6.7cm]{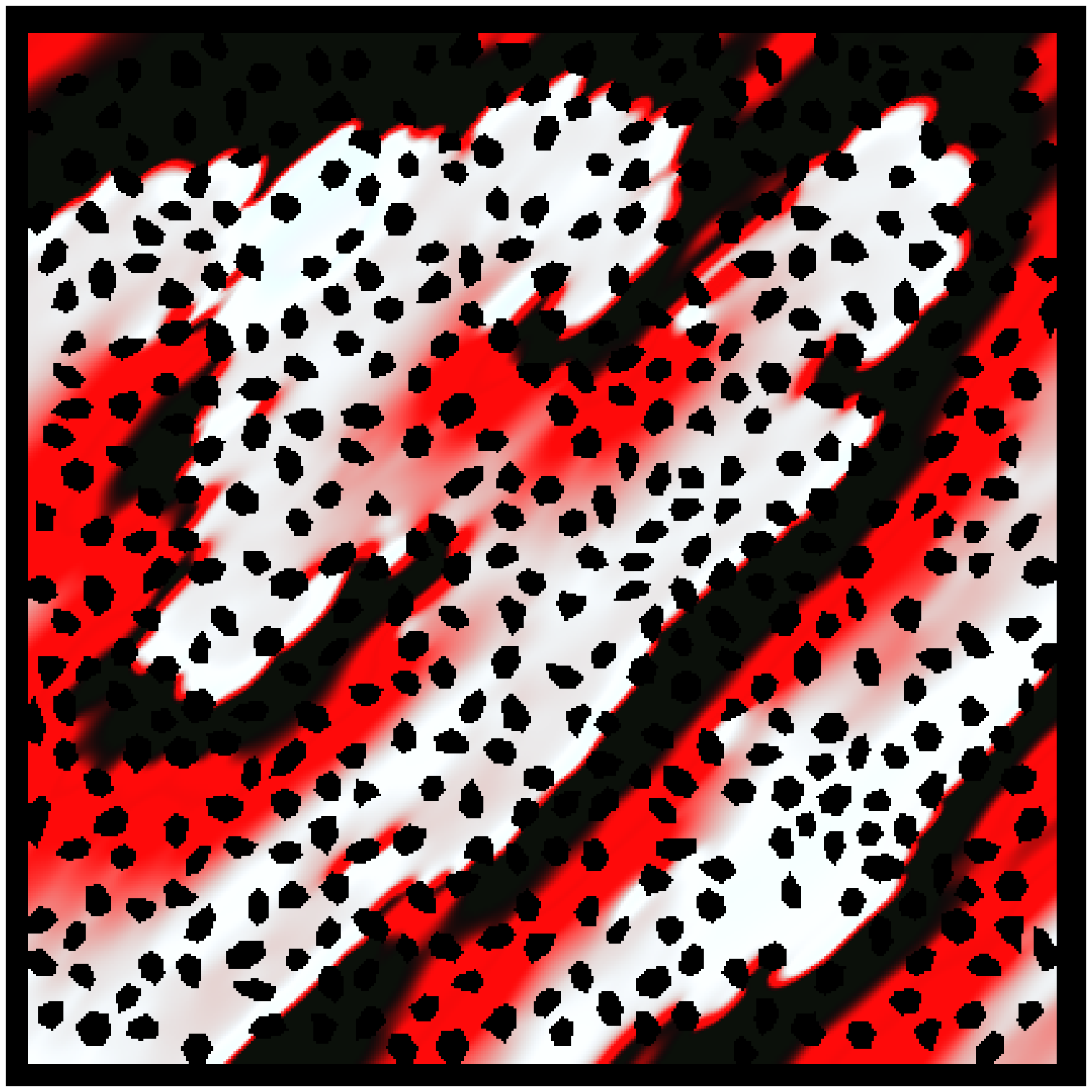}
  \caption{{\bf Effects of diffuse fibrosis.} Snapshots of spiral wave dynamics
   for ({\bf A}) control tissue and ({\bf B}) remodelled tissue (gap junctions,
   fibrosis) with a restitution slope of 1.4, taken at $t = 5000\,\mathrm{ms}$
   on a static domain. The patches of fibrotic tissue are clearly visible in the
   right hand panel and have an average area of $8.72\,\mathrm{mm}^2$ (mesh F0
   of Table \ref{tab:meshes1}).}
  \label{fig:UFibrosisPanel}
 \end{center}
\end{figure}

\begin{figure}
 \begin{center}
  \mbox{{\bf A} \hspace{6.7cm} {\bf B} \hspace{7.1cm}} \\ \vspace{-\baselineskip}
  \includegraphics[height=6.7cm]{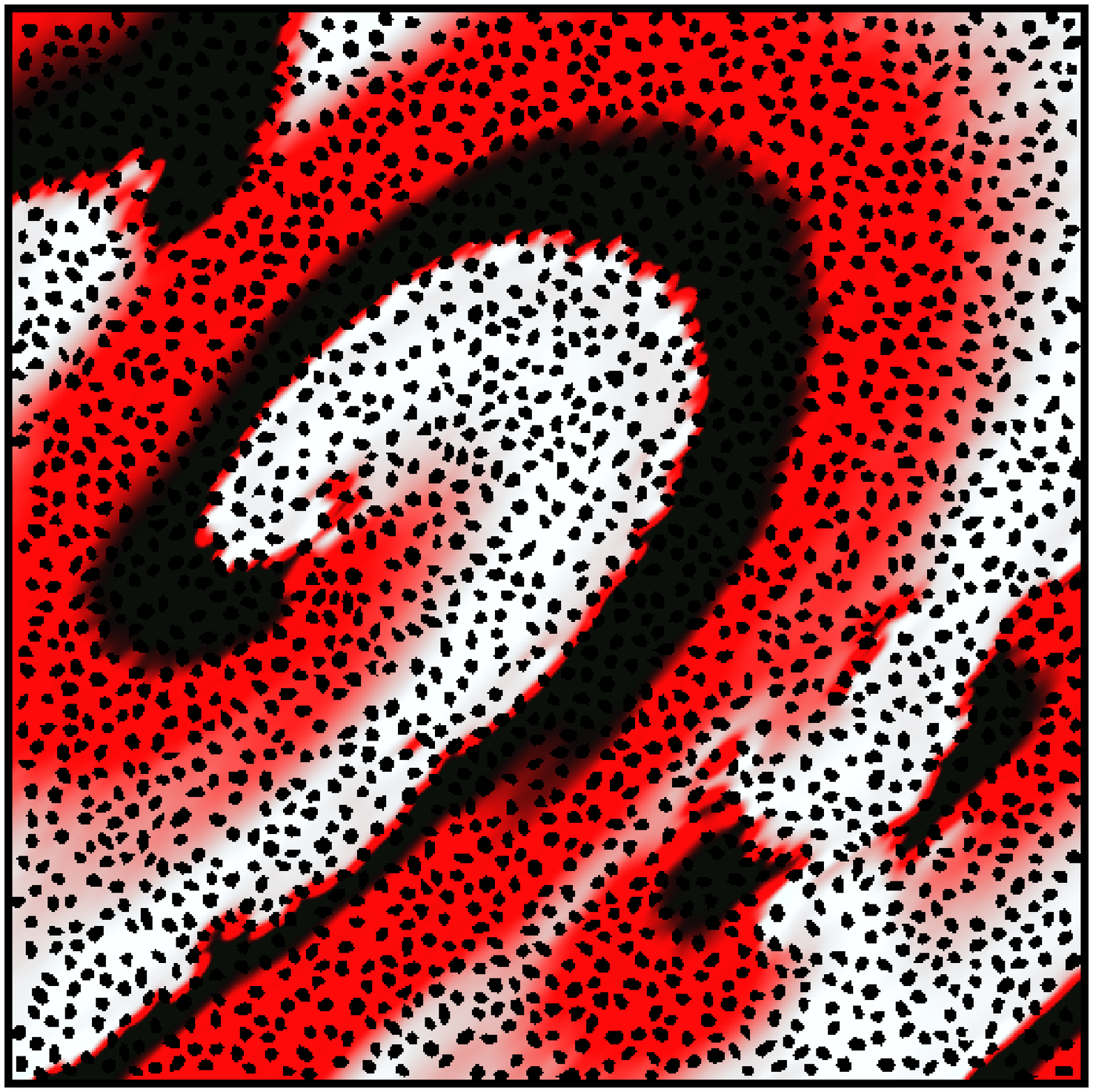} \quad
  \includegraphics[height=6.7cm]{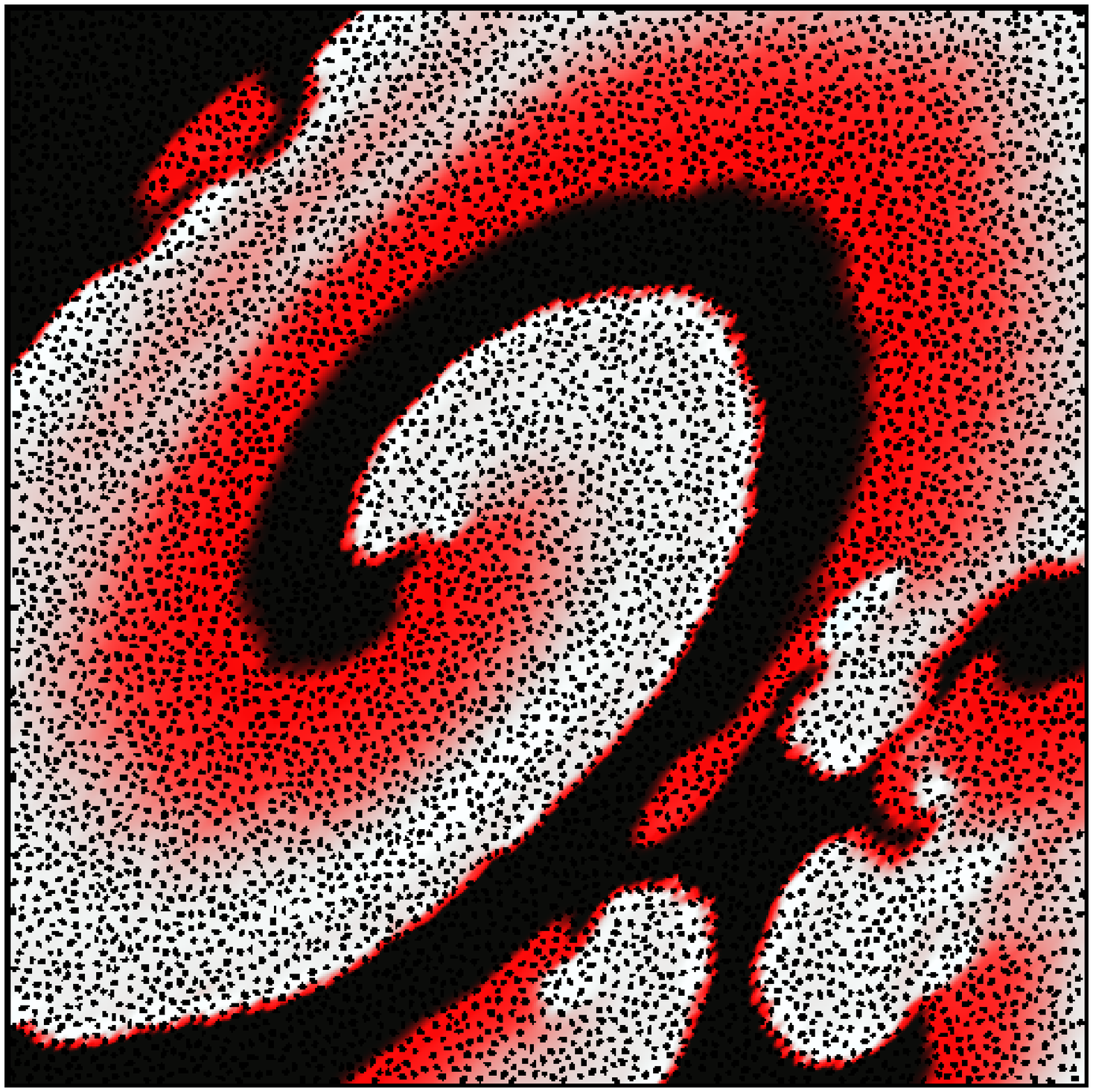} \\
  \vspace{4mm}
  \mbox{{\bf C} \hspace{7.0cm}} \\ \vspace{-\baselineskip}
  \includegraphics[height=6.7cm]{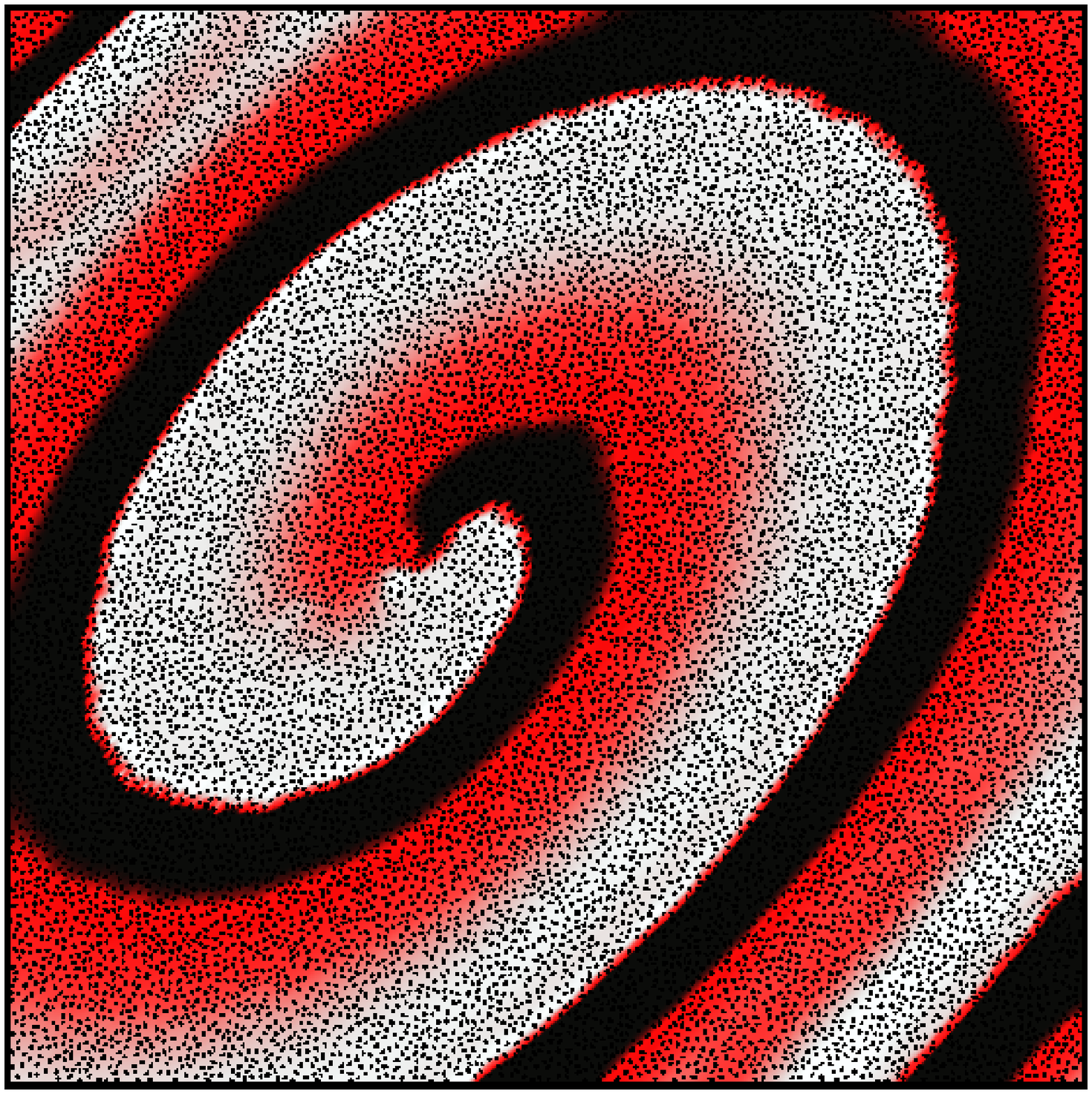}
  \caption{{\bf Effects of varying fibrotic patch size.} Snapshots of spiral wave
   dynamics for remodelled tissue (gap junctions, fibrosis) with a restitution
   slope of 1.4 on a static domain, sampled at ({\bf A}) $t = 13600\,\mathrm{ms}$
   on mesh F1, ({\bf B}) $t = 12360\,\mathrm{ms}$ on mesh F2 and ({\bf C})
   $t = 17400\,\mathrm{ms}$ on mesh F3 (all meshes from Table \ref{tab:meshes1}).
   In each case the fibrotic tissue constitutes approximately 27\% of the total
   domain area. The times for the snapshots are chosen to indicate the time that
   break-up first becomes clearly visible or, if break-up does not occur, the
   end-time of the numerical simulation.}
  \label{fig:Fib-2}
 \end{center}
\end{figure}

 Figure \ref{fig:UFibrosisPanel} shows that, in the case where a restitution slope
of 1.4 was used with mesh F0 (see Table \ref{tab:meshes1}), the mesh with the
largest fibrotic regions, the introduction of the remodelled tissue has caused an
otherwise stable spiral wave to break up. The spiral wave behaviour for the other
restitution slopes tested was not visibly affected: in particular, the effect was
not strong enough to destabilise the system when the restitution slope is 1.1. For
this reason, only the results obtained with a restitution slope of 1.4 have been
investigated further in this section. The results obtained when the domain was
allowed to deform are also not shown, since the mechanical deformation has a
further destabilising influence but not one strong enough to change the dynamics
significantly when the restitution slope was 1.1.
 The results presented in Figure \ref{fig:Fib-2} and summarised in the final row
of Table \ref{tab:meshes1} indicate that, as the individual fibrotic regions
become smaller, their destabilising effect is progressively reduced.

\begin{figure}
 \begin{center}
  \mbox{{\bf A} \hspace{6.7cm} {\bf B} \hspace{7.1cm}} \\ \vspace{-\baselineskip}
  \includegraphics[height=6.7cm]{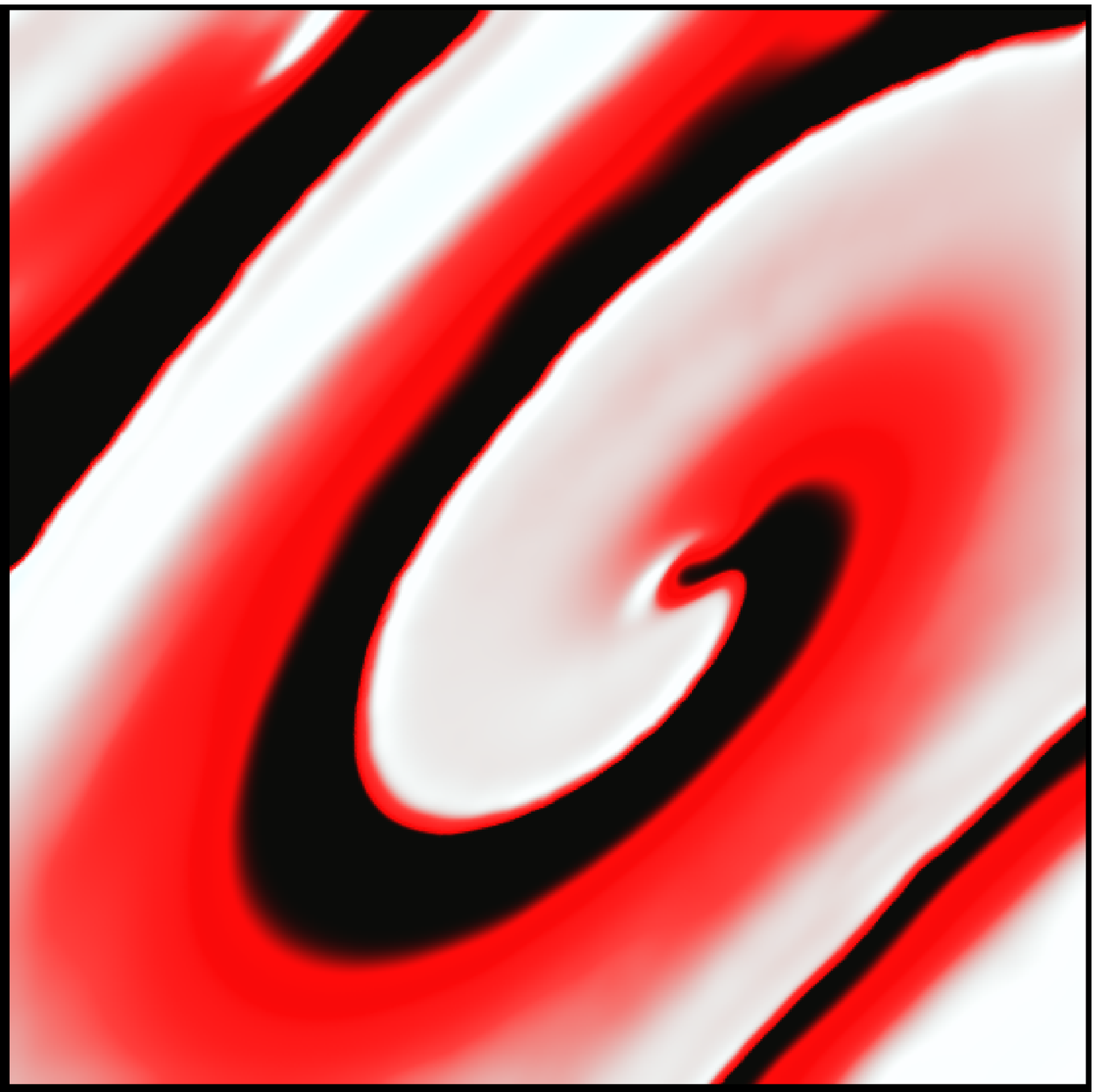} \quad
  \includegraphics[height=6.7cm]{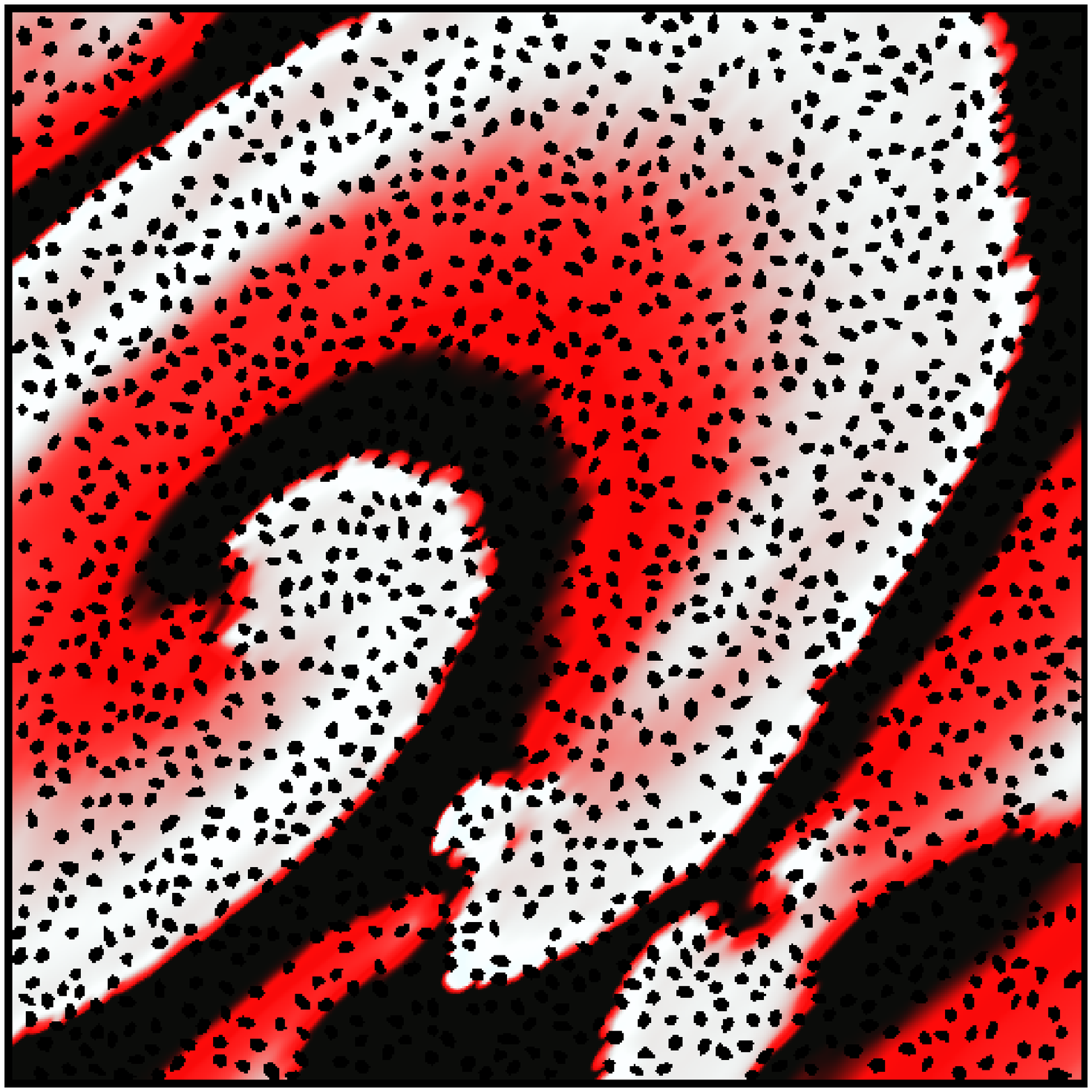} \\
  \vspace{4mm}
  \mbox{{\bf C} \hspace{6.7cm} {\bf D} \hspace{7.1cm}} \\ \vspace{-\baselineskip}
  \includegraphics[height=6.7cm]{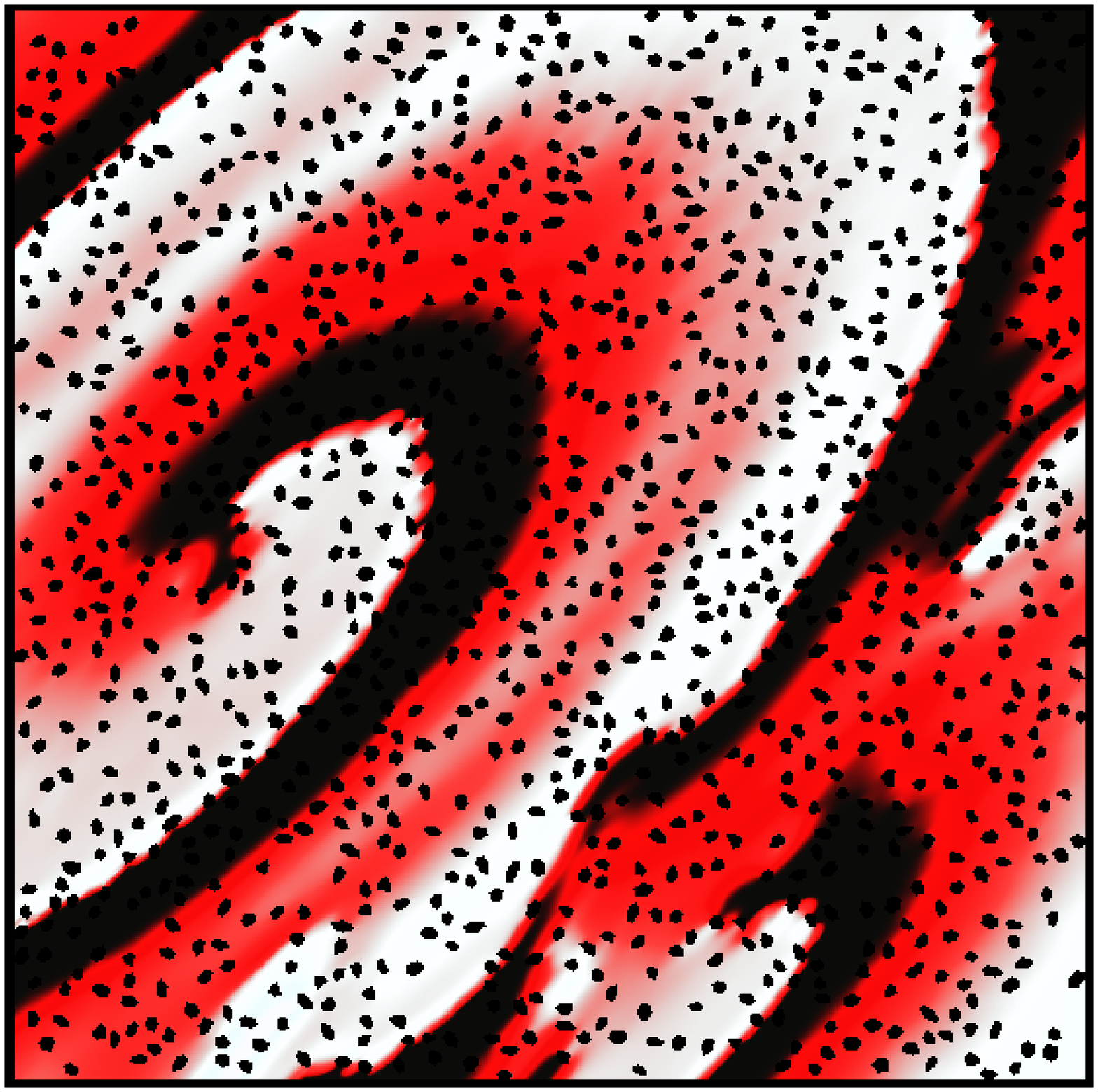} \quad
  \includegraphics[height=6.7cm]{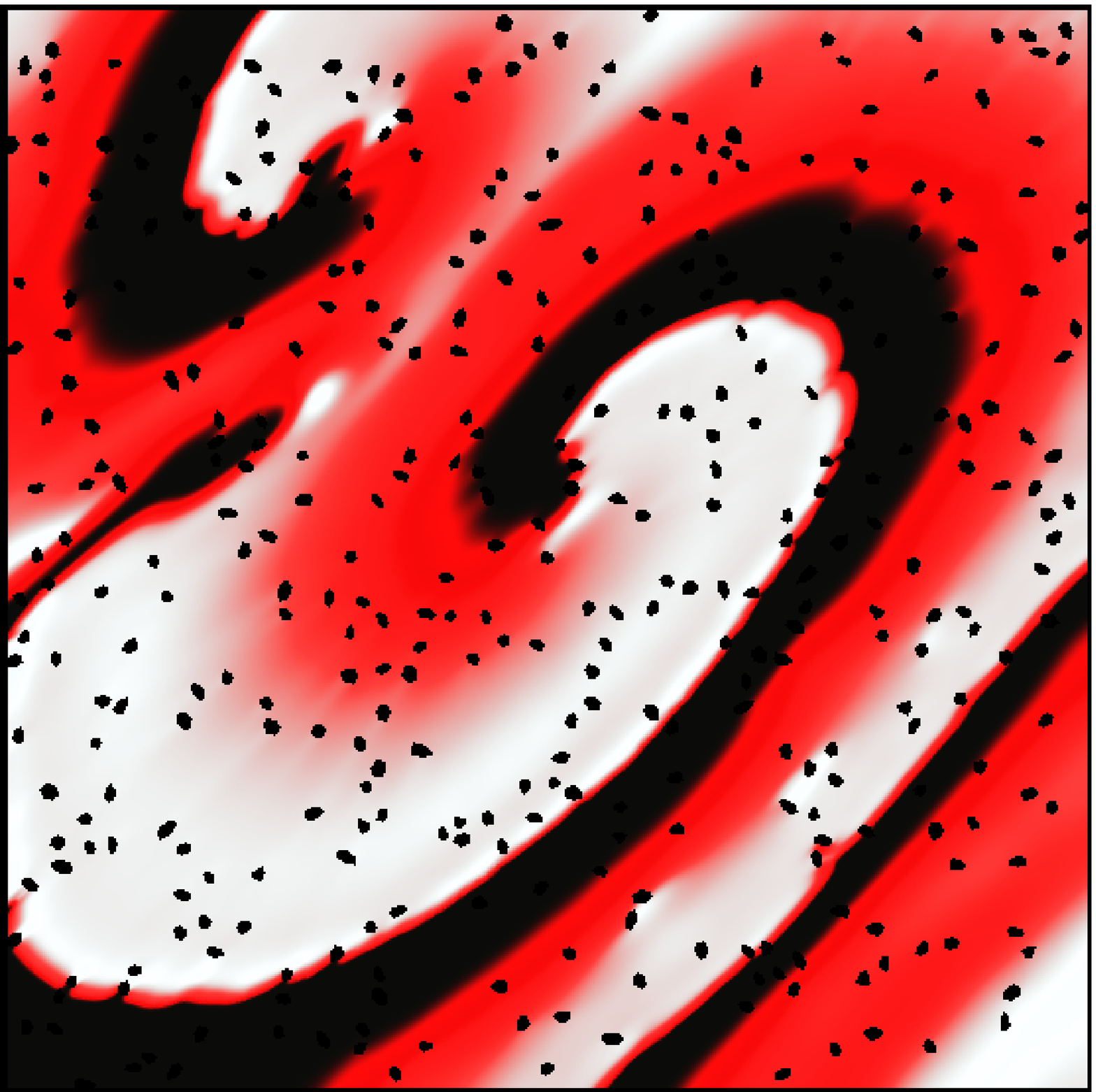}
  \caption{{\bf Effect of varying fibrotic patch density.} Snapshots of spiral
   wave dynamics for remodelled tissue (gap junctions, fibrosis) with a restitution
   slope of 1.4 on a static domain, sampled at ({\bf A}) $t = 10720\,\mathrm{ms}$
   on the control mesh N0, ({\bf B}) $t = 8600\,\mathrm{ms}$ on mesh F6, ({\bf C})
   $t = 8200\,\mathrm{ms}$ on mesh F7 and ({\bf D}) $t = 10720\,\mathrm{ms}$ on
   mesh F8 (all meshes from Table \ref{tab:FibAreasLower}). In each case the
   individual fibrotic regions are, on average, approximately $1.93\,\mathrm{mm}^2$
   in area. The times for the snapshots are chosen to indicate the time that
   break-up first becomes clearly visible or, if break-up does not occur, the
   end-time of the numerical simulation.}
  \label{fig:Fib-7}
 \end{center}
\end{figure}

 Next, the average area of the individual fibrotic regions was fixed at
$1.93\,\mathrm{mm}^2$ (mesh F1) and the proportion of the total tissue area
designated as fibrotic was varied, as indicated in Table \ref{tab:FibAreasLower}.
The range of fibrotic tissue density is kept within the bounds seen in real tissue
\cite{Rossi2001,TenTusscher2007a}. It can be seen in Figure \ref{fig:Fib-7} that,
even with smaller proportions of the domain designated as inexcitable, the spiral
wave is still destabilised. Furthermore, there is no clear dependence of the time
taken to the start of the spiral wave break-up on the proportion of fibrotic tissue.
Even a small number of fibrotic patches can have a significant impact on spiral
wave stability: for example, in panel {\bf D} of Figure \ref{fig:Fib-7} a smaller
spiral wave is visible in the top-left corner of the domain which has split from
the main wave.

\begin{table}
 \begin{center}
  \begin{tabular}{||l||c|c|c|c|c||} \hline                       
   Mesh                  &    F1    &    F5    &    F6    &    F7    &    F8    \\ \hline
   Inexcitable Region    &  26.81\% &  24.81\% &  20.17\% &  15.39\% &   5.09\% \\ \hline
   Spiral Wave Stability & Unstable & Unstable & Unstable & Unstable & Unstable \\ \hline
  \end{tabular}
  \caption{{\bf The computational domains used for the fibrosis simulations
   investigating the effects of varying the total proportion of fibrotic tissue.}
   The columns describe meshes in which the proportion of the domain designated
   as fibrotic (and hence non-conducting) is varied. In each case the average area
   of the individual fibrotic regions is chosen to be $1.93\,\mathrm{mm}^2$, as in
   mesh F1, repeated from Table \ref{tab:meshes1}. The final row indicates the
   stability of a spiral wave in a static domain when the restitution slope is
   1.4. All meshes produced an unstable spiral wave in a deforming domain.}
  \label{tab:FibAreasLower}
 \end{center}
\end{table}

\subsection*{Combining Electrophysiology and Tissue Remodelling}

\begin{figure}
 \begin{center}
  \mbox{{\bf A} \hspace{6.7cm} {\bf B} \hspace{7.1cm}} \\ \vspace{-\baselineskip}
  \label{fig:BaseC18-150}\includegraphics[width=6.7cm]{./BaseC18-150k.eps} \quad
  \label{fig:FibC18-150a}\includegraphics[width=6.7cm]{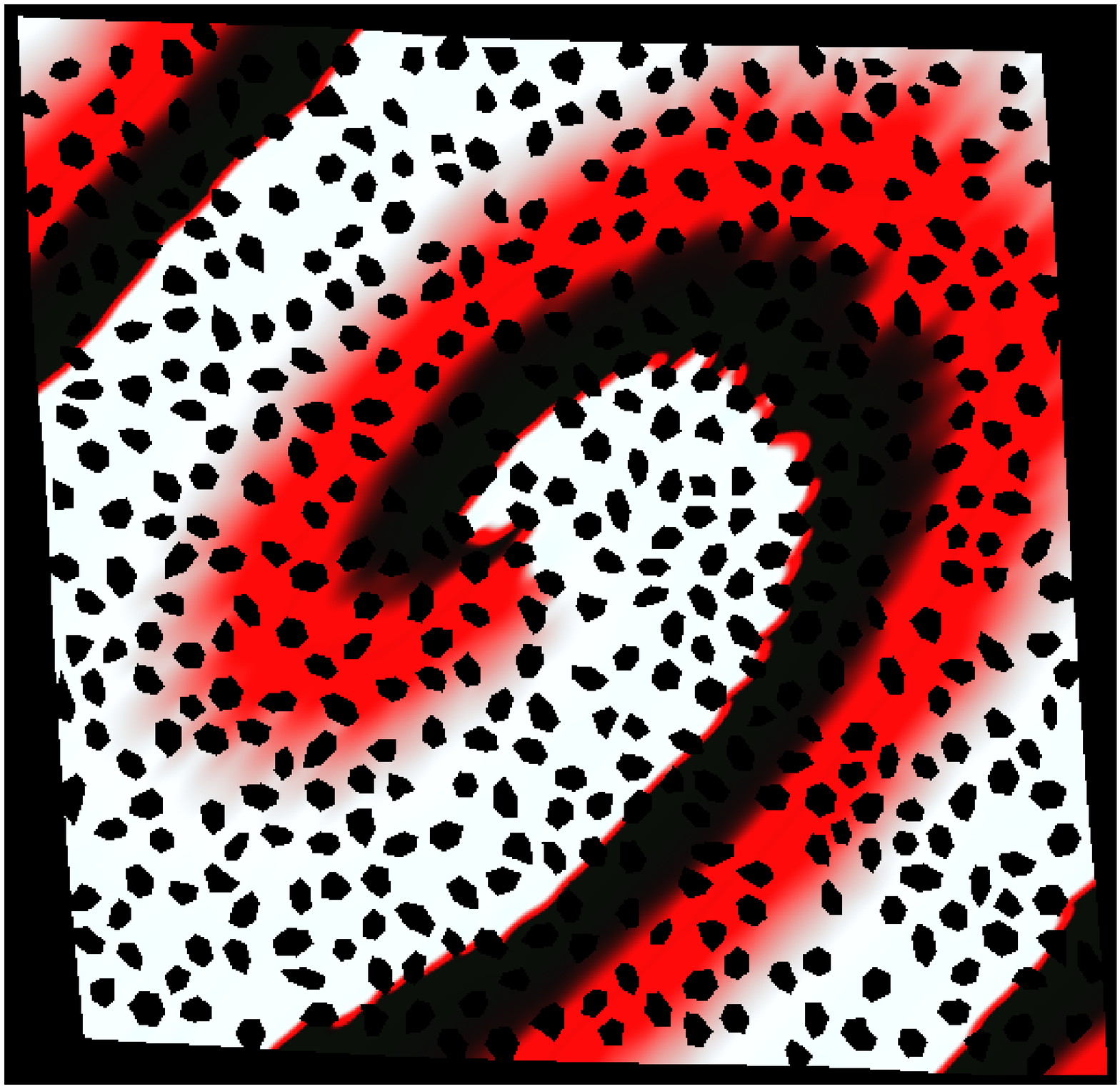}
  \caption{{\bf Effect of full electrophysiology and tissue remodelling.}
   Snapshots of spiral wave dynamics for ({\bf A}) control tissue on mesh
   N0 and ({\bf B}) remodelled electrophysiology and tissue (gap junctions, fibrosis)
   on mesh F0, with a restitution slope of 1.8, taken at $t = 7000\,\mathrm{ms}$
   on a static domain.}
  \label{fig:FibCuC}
 \end{center}
\end{figure}

 Finally, all of the proposed enhancements and modifications were included,
{\em i.e.}\ coupled electromechanics on a deforming domain, remodelled
electrophysiology, gap junction remodelling and regions of inexcitable fibrotic
tissue. Snapshots of the behaviour for a typical run are shown in Figure
\ref{fig:FibCuC}. Panel {\bf B} shows that in this case the spiral wave remains
stable for a restitution slope of 1.8. The combination of the restitution slope
of 1.8 and the deforming domain is the least stable configuration considered in
this work: all other combinations tested remained stable.

\section*{Discussion}

 In this paper, a computational model of coupled cardiac electromechanical
activity was described and used to investigate the effect on spiral wave stability
of changes in cardiac electrophysiology and tissue mechanics manifest in end-stage
heart failure. The electrophysiology was represented by the second-generation
cellular model proposed by ten Tusscher and Panfilov \cite{TenTusscher2006}
combined with anisotropic diffusion, the latter providing more rapid diffusion
along the fibre direction. The mechanical properties were based on those of an
incompressible Mooney-Rivlin material \cite{Mooney1940,Rivlin1948} in which, to
simulate anisotropy, the active tension acted only along the fibre direction. The
active tension, which governs the mechanical response to the electrical dynamics,
was estimated using a phenomenological dependence on voltage \cite{Nash2004}.
 Analogous experiments have been carried out \cite{Kirk2012} in which the active
tension was instead chosen to depend on the local calcium concentration, but this
led to the same conclusions about the stabilising and destabilising effects of
the remodelling and mechanical feedback.

 The finite element method was used to approximate both the electrophysiology and
mechanics models on unstructured triangular meshes, offering the flexibility to
model geometrically complex domains in the future. For both systems, numerical
methods were chosen for the spatial discretisation which did not require artificial
stabilisation techniques. A semi-implicit time-stepping scheme was used for the
electrophysiology to ensure that the length of the time-step allowed by stability
constraints on the computation was not limited (via the diffusion term) by the
resolution of the spatial mesh. The nature of the mathematical models and their
subsequent discretisation dictates that not only is the discrete mechanics system
about five times the size of the discrete electrophysiological system if
approximated on the same computational mesh, it is also highly nonlinear --
therefore it is the solution of the discrete mechanics system that dominates the
computation time. However, since the mechanical behaviour acts over longer length
scales than the electrical behaviour it does not need such a fine computational
mesh to resolve the dynamics so, in this work, the mechanics system was discretised
on a much coarser mesh. Interpolation between meshes was avoided by defining the
mesh on which the electrophysiology system was discretised to be a uniform
refinement of that on which the mechanics system was discretised. 

 The efficiency of the computational
algorithm was improved by the application of an ILU preconditioner
\cite{Saad1996,Chen2005} to the linear systems of equations which need to be solved
for each implicit time-step in the electrophysiology system and at each iteration
of the Newton solver applied to the nonlinear mechanics system. This is a
general-purpose preconditioner which, by careful choice of the algorithm
parameters, typically improved the computational performance by a factor of about
20. Preliminary results, presented in \cite{Kirk2012} but not shown here, suggest
that the efficiency could be further improved with the aid of adaptive mesh
refinement.
 Furthermore, this preconditioner is not optimal, and there is the potential to
construct a multigrid-based preconditioner for which the number of iterations
required to reach convergence is independent of the problem size (and hence the
mesh resolution).

 The spiral wave dynamics were investigated by studying three test cases,
distinguished by sets of conductance parameter values chosen to give different
dynamic restitution slopes (1.1, 1.4, 1.8) demonstrating progressively less stable
dynamics \cite{TenTusscher2006}. With the chosen parameters the spiral wave
generated by the TP06 model {\em without} mechanical deformation is stable for
restitution slopes of 1.1 and 1.4 but unstable for a restitution slope of 1.8.
When mechanical deformation is included the spiral wave becomes unstable for a
restitution slope of 1.4 (but remains stable for 1.1), as indicated in Table
\ref{tab:summary}. This supports the findings in \cite{Nash2004,Kirk2011}, that
including mechanical response and feedback can reduce the stability of a spiral
wave.

\begin{table}
 \begin{center}
  \begin{tabular}{|l|ccc|} \hline
   \multicolumn{4}{|c|}{Static Tissue}                                             \\ \hline
                                          & \multicolumn{3}{c|}{Restitution Slope} \\
                                          &     1.1     &     1.4     &    1.8     \\ \hline
   Healthy (Control) Tissue               &   Stable    &   Stable    & Unstable   \\
   Electrophysiology Remodelling Only     &   Stable    &   Stable    &  Stable    \\
   Tissue Remodelling Only                &   Stable    &  Unstable   & Unstable   \\
   Electrophysiology + Tissue Remodelling &   Stable    &   Stable    &  Stable    \\ \hline\hline
   \multicolumn{4}{|c|}{Deforming Tissue}                                          \\ \hline
                                          & \multicolumn{3}{c|}{Restitution Slope} \\
                                          &     1.1     &     1.4     &    1.8     \\ \hline
   Healthy (Control) Tissue               &   Stable    &  Unstable   & Unstable   \\
   Electrophysiology Remodelling Only     &   Stable    &   Stable    &  Stable    \\
   Tissue Remodelling Only                &   Stable    &  Unstable   & Unstable   \\
   Electrophysiology + Tissue Remodelling &   Stable    &   Stable    &  Stable    \\ \hline
  \end{tabular}
  \caption{{\bf Summary of spiral wave stability.} Each row represents a combination
   of static or deforming domain, and whether or not remodelling of electrophysiology
   and/or tissue (gap junctions, fibrosis) is introduced. The stability in the
   presence of diffuse fibrosis is assessed on the basis of numerical simulations
   carried out on mesh F0 of Table \ref{tab:meshes1}.}
  \label{tab:summary}
 \end{center}
\end{table}

\begin{figure}
 \begin{center}
  \mbox{{\bf A} \hspace{7.0cm} {\bf B} \hspace{6.8cm}} \\ \vspace{-\baselineskip}
  \label{fig:APDRest11}\includegraphics[width=7.5cm]{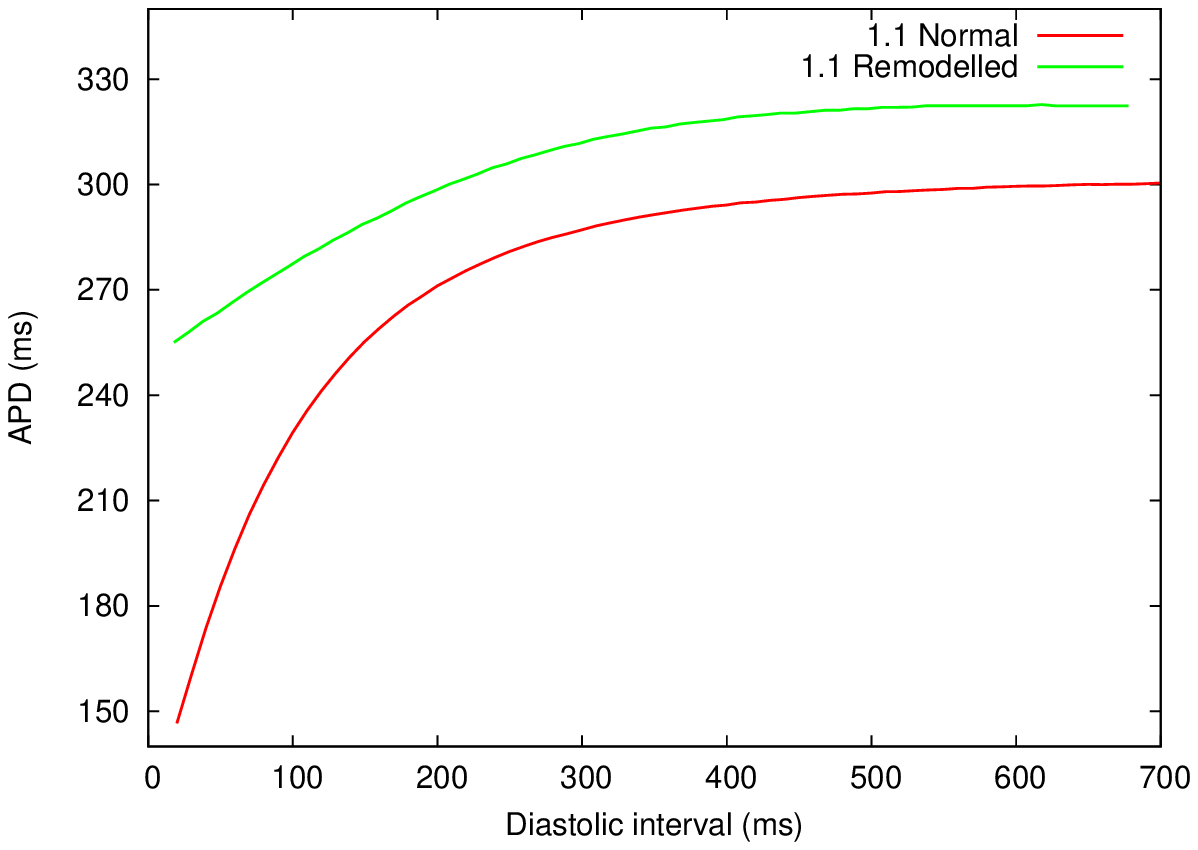}
  \label{fig:SingleCell1118}\includegraphics[width=7.5cm]{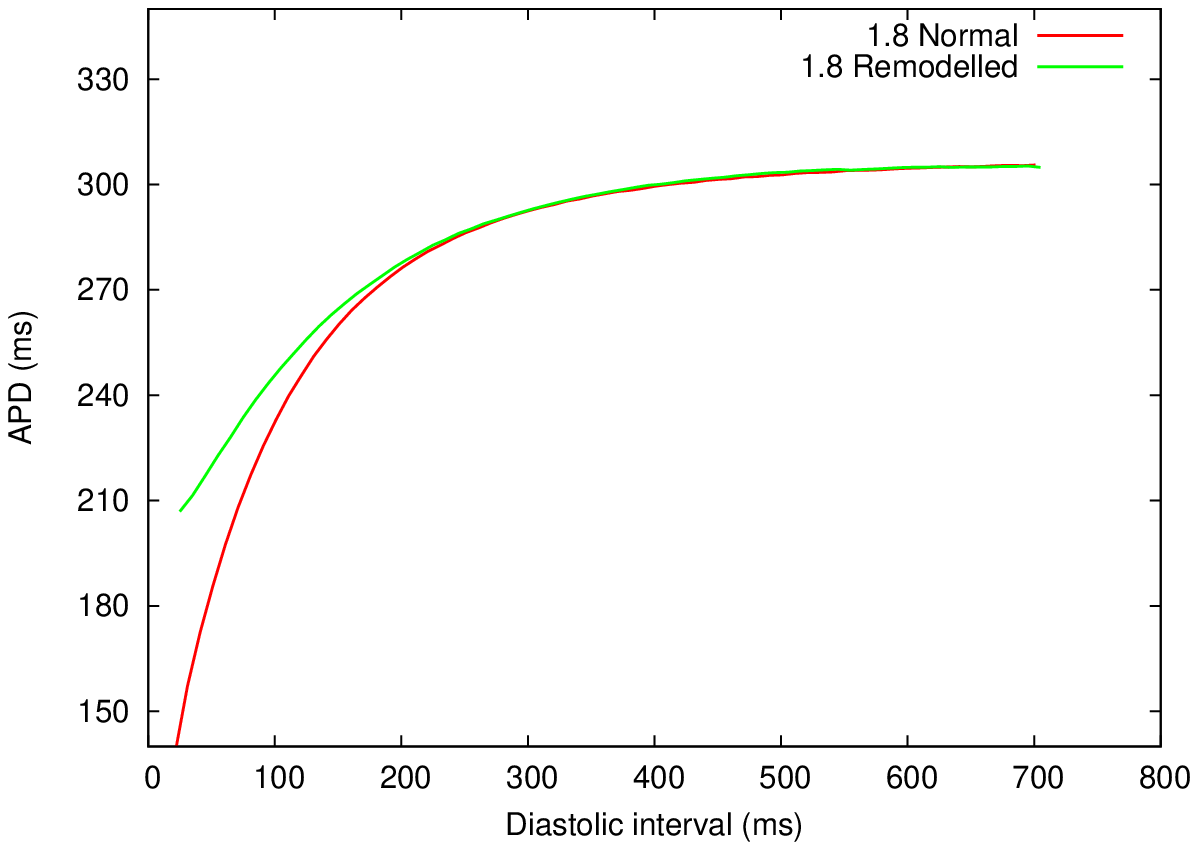}
  \caption{{\bf Remodelled restitution slopes.} Comparison of restitution slopes
   for normal and remodelled electrophysiology: ({\bf A}) 1.1; ({\bf B}) 1.8.}
  \label{fig:Rest2}
 \end{center}
\end{figure}

 Both the electrophysiology and tissue models were modified to imitate the effects
of end-stage heart failure. It was seen that electrophysiology remodelling alone,
achieved simply by modifying the parameters in the TP06 model \cite{TenTusscher2006},
can have a stabilising effect on spiral wave dynamics (see Figure \ref{fig:CurrC}
and Table \ref{tab:summary}).
 The reasons for this may be attributed to the effect that the electrophysiology
remodelling has on the restitution slope. It can be seen in Figure \ref{fig:Rest2}
that the remodelled electrophysiology produces a much shallower profile for the
restitution slope and the more stable configurations have already been seen to be
those with lower restitution slopes (see, for example, Figure \ref{fig:TP06-18}
or \cite{TenTusscher2006}).

 To simulate tissue fibrosis, gap junction remodelling was introduced by modifying
the diffusion tensor and patches of inexcitable tissue were created in the
computational mesh. We observed generally that this tissue remodelling can
destabilise spiral waves (see, for example, Figure \ref{fig:UFibrosisPanel}), and
investigated further the effect of varying the size and density of the patches of
fibrotic tissue.
 The results of these follow-on experiments suggest that reducing the size of the
individual regions of inexcitable tissue also weakens the observed destabilising
effect (Figure \ref{fig:Fib-2}).
 This contrasts with the observations of ten Tusscher and Panfilov \cite{TenTusscher2007a},
whose simulations suggest that diffuse fibrosis stabilises spiral waves supported
by the dynamic behaviour of the TP06 model. This might be attributable to the fact
that in that work, each fibrotic region was a $0.25\,\mathrm{mm} \times 0.25\,\mathrm{mm}$
square, its dimensions determined by the finite difference mesh they used. Each
region is therefore more uniform and approximately one quarter the area of the
smallest regions used in our study (Figure \ref{fig:Fib-2}), where the effects
of our remodelling is weakest. 

 The effect of changing the density of the fibrotic regions for a fixed patch size
was less pronounced. In fact, from our numerical simulations it was difficult to
determine a consistent pattern and it appears that the size of the fibrotic regions,
rather than the percentage of the domain that is inexcitable, is more significant
in determining the influence of fibrosis on spiral wave dynamics for the TP06 model.

 When the electrophysiology and tissue remodelling is combined with mechanical
deformation, the overall effect in our experiments was one of stabilisation (see
Figure \ref{fig:FibCuC}). The stabilising effects of the electrophysiology
remodelling dominated the combined destabilising effects of tissue remodelling
and mechanical deformation. In fact, as summarised in Table \ref{tab:summary},
the simulations were always stable whenever the remodelled electrophysiology was
used.
 We suggest that this seemingly contradictory finding may be due to a lack of
spatial and/or temporal electrophysiological heterogeneity in the model. Such
heterogeneities, which are well known substrates for the initiation of re-entrant
arrhythmias \cite{Tomaselli1999,Lou2012}, may also play a role in the
destabilisation of re-entry and cause breakdown of re-entrant waves ({\em i.e.}\
degeneration from ventricular tachycardia to fibrillation) \cite{Keldermann2008}.
Including such heterogeneities in the model may increase the incidence of spiral
wave break-up, despite the stabilising effects of the cellular-level
electrophysiological remodelling, and their inclusion is therefore an important
next step in model development. Furthermore, although we have included gap junction
remodelling in our study, our model does not include a description of the fibre
disruption that is commonly seen in heart failure and which has been implicated
in arrhythmogenesis (e.g. \cite{Benoist2012}). Whether or not this particular
type of fibre disruption can also result in destabilisation of re-entrant waves
remains to be investigated in detail, although it does appear likely
\cite{Benson2011b}. Nevertheless, and despite these two potential areas for
further study, our model has highlighted several electromechanical mechanisms
that increase or decrease the propensity of ventricular tachycardia to degenerate
into ventricular fibrillation.

\vspace{\baselineskip}
 In this paper, the TP06 model of electrophysiology has been coupled with a
Mooney-Rivlin model of mechanical response, and additional assumptions (outlined
in the paper) have been made about the remodelling effect that failing tissue
has on the dynamics of the system and the model parameters. For this combination
of models, we have observed that:
\begin{itemize}
 \item changes in electrophysiology introduced to simulate failing cardiac tissue
       can stabilise the electrical dynamics;
 \item the introduction of inexcitable fibrotic regions can cause a previously
       stable spiral wave to break up; 
 \item in such cases, the sizes of the individual fibrotic regions seem to have
       a stronger influence on stability than their overall density -- decreasing
       the size of the fibrotic regions increased stability;
 \item the stabilisation effects of the remodelled electrophysiology are strong
       enough to overcome break-up caused by the combined effects of mechanical
       deformation and fibrosis.
\end{itemize}

 These observations relate to a specific combination of models for simulating
coupled electromechanical behaviour, and it would be interesting for future
research to investigate whether they are indicative of the effects of diffuse
fibrosis on spiral wave stability for a broader range of models. The computational
framework presented here is flexible enough to provide the basis for a study of
this type, but improvements would need to be made to the efficiency of the underlying
numerical algorithms if simulations are to be carried out in three space dimensions
with meshes fine enough to resolve the wave-fronts generated by electrophysiology
models such as TP06. This is the subject of ongoing research, early results of
which indicate that mesh adaptivity can significantly improve computational
efficiency \cite{Kirk2012}. Furthermore, the replacement of the general-purpose
ILU preconditioner with an optimal (multigrid-based) preconditioner would remove the dependence of
the convergence of the iterative solver on user-defined parameters. If an efficient,
parallel, implementation of these techniques can be developed then the effects of
diffuse fibrosis can be explored more thoroughly and, ultimately, three-dimensional
simulations of cardiac dynamics in heterogeneous tissue could be carried out on
realistic geometries.

\section*{Acknowledgments}

 NRK was supported by an Engineering and Physical Sciences Research Council
Doctoral Training Grant. APB was supported by a Medical Research Council Special
Training Fellowship in Biomedical Informatics (G0701776).

\bibliography{BibNK}

\end{document}